\documentclass[12pt,a4paper]{article}

\usepackage{amsmath}
\usepackage{mathtools}
\usepackage{amsfonts}
\usepackage{amssymb}
\usepackage{graphicx}
\usepackage{color}
\usepackage{booktabs}
\usepackage{inputenc}
\usepackage[T1]{fontenc}
\usepackage{mathrsfs}
\usepackage{enumerate}
\usepackage{xcolor,url}
\usepackage{cancel,dsfont} 
\usepackage{multirow}
\usepackage[normalem]{ulem} 
\usepackage{makecell}
\usepackage{rotating}
\usepackage{cancel}

\newcommand{\eea}{\end{eqnarray}}
\newcommand{\bea}{\begin{eqnarray}}
\newcommand{\be}{\begin{equation}}
\newcommand{\ee}{\end{equation}}

\newcommand{\kvec}{\vec{k}}
\newcommand{\qvec}{\vec{q}}
\newcommand{\rmd}{\mathrm{d}}
\newcommand{\calP}{\mathcal{P}}

\newcommand{\fnl}{f_{\rm NL}}
\newcommand{\xvec}{\vec{x}}
\newcommand{\vx}{\vec{x}}

\def\be{\begin{equation}}
\def\ee{\end{equation}}

\def\nn{\nonumber}

\def\knl{k_{\rm NL}}
\renewcommand{\(}{\left(}
\renewcommand{\)}{\right)}

\newcommand{\al}{\alpha}
\newcommand{\bt}{\beta}
\newcommand{\eps}{\epsilon}

\def\hinvMpc{h\,{\rm Mpc}^{-1}}
\def\Mpcinvh{{\rm Mpc}/h}

\newcommand{\andd}{\ , \quad \text{and}  \quad}

\newcommand{\eqn}[1]{eq.~(\ref{#1})}

\newcommand{\secref}[1]{sec.~\ref{#1}}

\newcommand{\appref}[1]{app.~\ref{#1}}

\newcommand{\deltaone}{\delta^{(1)}}
\newcommand{\half}{{\textstyle{\frac12}}}

\newcommand{\kmax}{k_{\rm max }}

\newcommand{\code}[1]{\texttt{#1}}

\usepackage[colorlinks,bookmarks]{hyperref}
\definecolor{linkblue}{rgb}{0,0,0.8}
\definecolor{linkgreen}{rgb}{0,0.5,0}

\hypersetup{pdfpagemode=UseNone, pdfstartview=FitH, linkcolor=linkblue,
            citecolor=linkgreen, urlcolor=linkblue}

\begin{document}

\vspace*{-25mm}

\begin{flushright}
{\small NUHEP-TH/22-05}
\end{flushright}

\begin{center}

{\Large \bf The BOSS bispectrum analysis at one loop\\[0.3cm] {from the Effective Field Theory of Large-Scale Structure}}  \\[0.7cm]
{\large   Guido D'Amico${}^{1,2}$,  Yaniv Donath${}^{3}$, Matthew Lewandowski${}^{4}$,\\[0.3cm] Leonardo Senatore${}^{5}$, and  Pierre Zhang${}^{6,7,8}$ \\[0.7cm]}

\end{center}

%matt
\begin{center}

\vspace{.0cm}

\begin{small}

{ { \sl $^{1}$ Department of Mathematical, Physical and Computer Sciences,\\ University of Parma, 43124 Parma, Italy}}
\vspace{.05in}

{ { \sl $^{2}$ INFN Gruppo Collegato di Parma, 43124 Parma, Italy}}
\vspace{.05in}

{ { \sl $^{3}$ Department of Applied Mathematics and Theoretical Physics,\\
University of Cambridge, Cambridge, CB3 OWA, UK}}
\vspace{.05in}

{ { \sl $^{4}$ Department of Physics and Astronomy,\\ Northwestern University, Evanston, IL 60208}}
\vspace{.05in}

{ { \sl $^{5}$ Institut fur Theoretische Physik, ETH Zurich,
8093 Zurich, Switzerland}}
\vspace{.05in}

{ { \sl $^{6}$ Department of Astronomy, School of Physical Sciences, \\
University of Science and Technology of China, Hefei, Anhui 230026, China}}
\vspace{.05in}

{ { \sl $^{7}$ CAS Key Laboratory for Research in Galaxies and Cosmology, \\
University of Science and Technology of China, Hefei, Anhui 230026, China}}
\vspace{.05in}

{ { \sl $^{8}$ School of Astronomy and Space Science, \\
University of Science and Technology of China, Hefei, Anhui 230026, China}}
\vspace{.05in}

\end{small}
\end{center}

\hrule \vspace{0.3cm}
{\small  \noindent \textbf{Abstract} 
  \noindent  We analyze the BOSS power spectrum monopole and quadrupole, and the bispectrum monopole and quadrupole data, using the predictions from the Effective Field Theory of Large-Scale Structure (EFTofLSS). Specifically, we use the one loop prediction for the power spectrum and the bispectrum monopole, and the tree level for the bispectrum quadrupole. After validating our pipeline against numerical simulations as well as checking for several internal consistencies, we apply it to the observational data. We find that analyzing the bispectrum monopole to higher wavenumbers thanks to the one-loop prediction, as well as the addition of the tree-level quadrupole, significantly reduces the error bars with respect to our original analysis of the power spectrum at one loop {and bispectrum monopole at tree level}. After fixing the spectral tilt to Planck preferred value and using a Big Bang Nucleosynthesis prior, we measure $\sigma_8=0.794\pm 0.037$, $h = 0.692\pm 0.011$, and $\Omega_m = 0.311\pm 0.010$ to about $4.7\%$, $1.6\%$, and $3.2\%$, at $68\%$ CL, respectively. 
This represents an error bar reduction with respect to the power spectrum-only analysis of about $30\%$, $18\%$, and $13\%$ respectively.
 Remarkably, the results are compatible with the ones obtained with a power-spectrum-only analysis, showing the power of the EFTofLSS in simultaneously predicting several observables. We find no tension with Planck.

\vspace{0.3cm}}
\hrule

\vspace{0.3cm}
\newpage

\tableofcontents

\section{Introduction, Main Results and Conclusion}  \label{sec:intro}

The SDSS-III Baryon Oscillation Spectroscopic Survey (BOSS) has mapped the clustering of galaxies in the nearby Universe in an unprecedented amount and with great accuracy~\cite{BOSS:2016wmc}. 
Although BOSS' survey volume is modest with respect to upcoming experiments such as DESI~\cite{2013arXiv1308.0847L} or Euclid~\cite{Amendola:2012ys}, the BOSS data are remarkable as they have been revealing a wealth of cosmological information from the large-scale structure of the Universe.   
\vspace{0.5cm}

In the last couple of years, the Effective Field Theory of Large-Scale Structure~(EFTofLSS) prediction at one-loop order has been used to analyze the BOSS Full Shape (FS) of the {galaxy} Power Spectrum (PS)~\cite{DAmico:2019fhj,Ivanov:2019pdj,Colas:2019ret}, and Correlation Function~(CF)~\cite{Zhang:2021yna,Chen:2021wdi}. 
The BOSS galaxy-clustering bispectrum monopole using the tree-level prediction was first analyzed in~\cite{DAmico:2019fhj}   (see~\cite{Philcox:2021kcw} for a recent slight generalization). {See also \cite{Sugiyama:2018yzo, Sugiyama:2019ike, Sugiyama:2020uil} for other techniques and analysis using linear theory with higher multipoles.}  All $\Lambda$CDM cosmological parameters have been measured from these data by only imposing a prior from Big Bang Nucleosynthesis {(BBN)}, reaching a remarkable, and perhaps surprising, precision on some of these. For example, the present amount of matter, $\Omega_m$, and the Hubble constant (see also~\cite{Philcox:2020vvt,DAmico:2020kxu} for subsequent refinements) have error bars that are not far from the ones obtained from the Cosmic Microwave Background (CMB)~\cite{Planck:2018vyg}.
For clustering and smooth quintessence models, limits on the dark energy equation of state $w$ parameter of $\lesssim 5\%$ have been set using only late-time measurements~\cite{DAmico:2020kxu,DAmico:2020tty}. This is again quite close to the ones obtained with the CMB~\cite{Planck:2018vyg}.  These measurements provide a new, CMB-independent, method for determining the Hubble constant~\cite{DAmico:2019fhj}, resulting in a measurement that is comparable, if not better, to the one based on the cosmic ladder~\cite{Riess:2019cxk,Freedman:2019jwv} and CMB.
Therefore, this tool has been used to shed light on how some models that were proposed to alleviate the tension in the Hubble measurements (see e.g.~\cite{Verde:2019ivm}) between the CMB and cosmic ladder~\cite{DAmico:2020ods,Ivanov:2020ril} (see also~\cite{Niedermann:2020qbw,Smith:2020rxx}) actually perform. 

Very recently, in~\cite{DAmico:2022gki}, we used the one-loop EFTofLSS prediction for the bispectrum to set the first and strong limits on primordial inflationary non-Gaussianities from Large-Scale Structure (LSS) (see also~\cite{Cabass:2022wjy,Cabass:2022ymb}  for a contemporary and a subsequent paper, where, once put together, the same shapes are constrained but stopping at the tree-level EFTofLSS prediction, and so obtaining much weaker constraints for the same data).  
 We obtained limits on three of the so-called $\fnl$ parameters, {$\fnl^{\rm equil.}= 217 \pm 297\,,\ \fnl^{\rm orth.}= -64 \pm 74\,,\ \fnl^{\rm loc.}= 49 \pm 36$}, at $68\%$ {confidence level}, which are predicted to be produced by some single-clock~\cite{Creminelli:2005hu,Senatore:2009gt} or multiple fields~\cite{{Bernardeau:2002jy,Lyth:2002my,Zaldarriaga:2003my,Babich:2004gb,Senatore:2010wk}} inflationary models.
{Perhaps quite surprisingly, those constraints were} already quite on par with the ones of the powerful CMB experiment WMAP~\cite{WMAP:2012fli}, though largely inferior to the more recent CMB experiment Planck~\cite{Planck:2019kim}. Significant limits from LSS on just $\fnl^{\rm loc.}$ were obtained using the power spectrum {only}, first in~\cite{Slosar:2008hx}, using the so-called non-local bias~\cite{Dalal:2007cu,Verde:2009hy,Schmidt:2010gw}, but the analysis of~\cite{DAmico:2022gki} uses for the first time the bispectrum, obtaining much stronger constraints using the data from the same experiments. \vspace{0.5cm}

It took an intense and years-long line of study to develop the EFTofLSS from the initial formulation to the level {that allows it} to be applied to data. It appears to us that it often happens that there is no proper acknowledgment of the many works that were needed to reach this point. For instance, several authors cite Refs.~\cite{DAmico:2019fhj,Ivanov:2019pdj} for the `model' to analyze the PS FS, but the EFT model that is used in~\cite{DAmico:2019fhj,Ivanov:2019pdj} is essentially the same as the one originally proposed in~\cite{Perko:2016puo}.  We therefore find it fair to add the following footnote in every paper where the EFTofLSS is used to analyze observational data.  
Even though some of the mentioned papers are not strictly required to analyze the data, we, and we believe probably {anybody} else, would {not} have applied the EFTofLSS to data without all these intermediate results that allowed us to overcome the widespread skepticism about the usefulness of the EFTofLSS.\footnote{The initial formulation of the EFTofLSS was performed in Eulerian space in~\cite{Baumann:2010tm,Carrasco:2012cv}, and subsequently extended to Lagrangian space in~\cite{Porto:2013qua}.
The dark matter power spectrum has been computed at one-, two- and three-loop orders in~\cite{Carrasco:2012cv, Carrasco:2013sva, Carrasco:2013mua, Carroll:2013oxa, Senatore:2014via, Baldauf:2015zga, Foreman:2015lca, Baldauf:2015aha, Cataneo:2016suz, Lewandowski:2017kes,Konstandin:2019bay}.
These calculations were accompanied by some  theoretical developments of the EFTofLSS, such as a careful understanding of renormalization~\cite{Carrasco:2012cv,Pajer:2013jj,Abolhasani:2015mra} (including rather-subtle aspects such as lattice-running~\cite{Carrasco:2012cv} and a better understanding of the velocity field~\cite{Carrasco:2013sva,Mercolli:2013bsa}), of several ways for extracting the value of the counterterms from simulations~\cite{Carrasco:2012cv,McQuinn:2015tva}, and of the non-locality in time of the EFTofLSS~\cite{Carrasco:2013sva, Carroll:2013oxa,Senatore:2014eva}.
These theoretical explorations also include an enlightening study in 1+1 dimensions~\cite{McQuinn:2015tva}.
An IR-resummation of the long displacement fields had to be performed in order to reproduce the Baryon Acoustic Oscillation (BAO) peak, giving rise to the so-called IR-Resummed EFTofLSS~\cite{Senatore:2014vja,Baldauf:2015xfa,Senatore:2017pbn,Lewandowski:2018ywf,Blas:2016sfa}. 
Accounts of baryonic effects were presented in~\cite{Lewandowski:2014rca,Braganca:2020nhv}. The dark-matter bispectrum has been computed at one-loop in~\cite{Angulo:2014tfa, Baldauf:2014qfa}, the one-loop trispectrum in~\cite{Bertolini:2016bmt}, and the displacement field in~\cite{Baldauf:2015tla}.
The lensing power spectrum has been computed at two loops in~\cite{Foreman:2015uva}.
Biased tracers, such as halos and galaxies, have been studied in the context of the EFTofLSS in~\cite{ Senatore:2014eva, Mirbabayi:2014zca, Angulo:2015eqa, Fujita:2016dne, Perko:2016puo, Nadler:2017qto,Donath:2020abv} (see also~\cite{McDonald:2009dh}), the halo and matter power spectra and bispectra (including all cross correlations) in~\cite{Senatore:2014eva, Angulo:2015eqa}. Redshift space distortions have been developed in~\cite{Senatore:2014vja, Lewandowski:2015ziq,Perko:2016puo}. 
Neutrinos have been included in the EFTofLSS in~\cite{Senatore:2017hyk,deBelsunce:2018xtd}, clustering dark energy in~\cite{Lewandowski:2016yce,Lewandowski:2017kes,Cusin:2017wjg,Bose:2018orj}, and primordial non-Gaussianities in~\cite{Angulo:2015eqa, Assassi:2015jqa, Assassi:2015fma, Bertolini:2015fya, Lewandowski:2015ziq, Bertolini:2016hxg}.
Faster evaluation schemes for the calculation of some of the loop integrals have been developed in~\cite{Simonovic:2017mhp}.
Comparison with high-quality $N$-body simulations to show that the EFTofLSS can accurately recover the cosmological parameters have been performed in~\cite{DAmico:2019fhj,Colas:2019ret,Nishimichi:2020tvu,Chen:2020zjt}.} \vspace{0.5cm}

In this paper we upgrade our original analysis of the one-loop power spectrum monopole {and quadrupole and tree-level bispectrum monopole~\cite{DAmico:2019fhj}, to include the full one-loop bispectrum monopole and the tree-level bispectrum quadrupole.} 
We scan over all the $\Lambda$CDM parameters with BBN prior on the baryon abundance, $\Omega_b h^2$, with the exception of the tilt, $n_s$, that we fix to the Planck preferred value. 

The development of a pipeline that allows us to analyze the one-loop bispectrum predicted by the EFTofLSS  has required much theoretical work, and the explanation  of such techniques will be presented in {two} upcoming papers~\cite{DAmico:2022ukl, Anastasiou:2022udy}. While the application to constrain primordial non-Gaussianities was already presented in~\cite{DAmico:2022gki}, here, instead, we will just give the essential details and focus on constraints of the $\Lambda$CDM parameters.  

\begin{figure}[t!]
  \centering
  \includegraphics[width=0.67\textwidth]{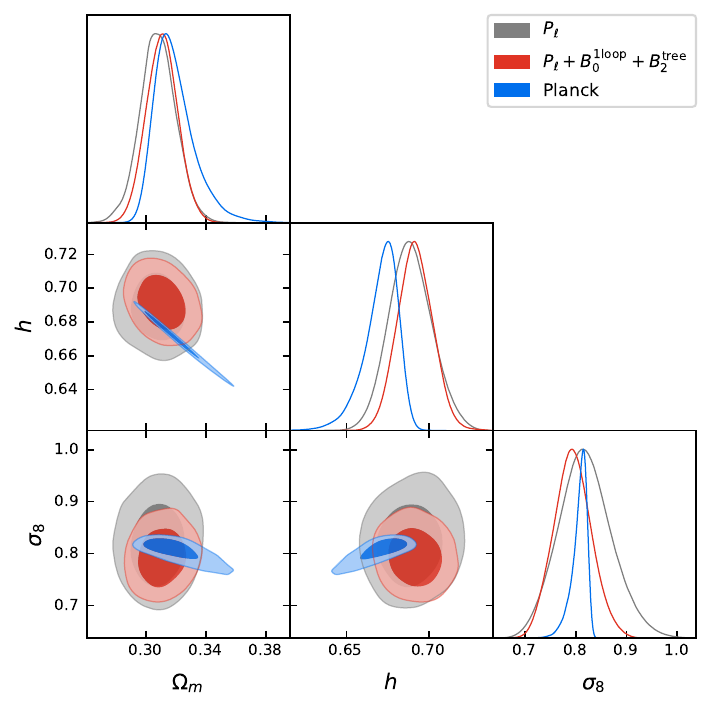}
    \vspace{0.5em}
    
\scriptsize
    \begin{tabular}{|l|c|c|c|c|c|c|c|}
     \hline 
    \makecell{\textrm{best-fit}\\$\text{\rm mean} \pm \sigma$} 		& $\Omega_m$ 			 &  $ h$					&  $\sigma_8 $ 					& $\omega_{cdm}$ 				& $\ln (10^{10}A_s)$ 	& $S_8$	\\ \hline 
    $P_\ell$ 					&  \makecell{0.2984 \\ $0.308\pm 0.012$}			 & \makecell{0.6763 \\ $0.689^{+0.012}_{-0.014}$} 	& \makecell{0.8305 \\ $0.819^{+0.049}_{-0.055}$} 		& \makecell{0.1143 \\ $0.1232\pm 0.0075$} 			& \makecell{3.123 \\ $3.02\pm 0.15$} 		& \makecell{0.8283 \\ $0.830^{+0.051}_{-0.060}$}		\\ \hline
        $P_\ell + B_0^\mathrm{tree}$ 				&  \makecell{0.3101\\$0.309\pm 0.011$}			 & \makecell{ 0.6907\\$0.691\pm 0.012$} 		& \makecell{0.8063\\$0.804\pm 0.049$	}	& \makecell{0.1248\\$0.1246\pm 0.0058$} 			& \makecell{2.98\\$2.97\pm 0.13$} 		& \makecell{0.8197\\$0.816^{+0.050}_{-0.057}$}	\\ \hline
    $P_\ell + B_0^\mathrm{1loop}$ 				&  \makecell{0.3210\\$0.314\pm 0.011$}			 & \makecell{0.6956\\$0.693\pm 0.011$} 		& \makecell{0.7882\\$0.790^{+0.033}_{-0.037}$	}	& \makecell{0.1331\\$0.1278\pm 0.0061$} 			& \makecell{2.82\\$2.90\pm 0.11$} 		& \makecell{0.8153\\$0.807^{+0.037}_{-0.043}$}	\\ \hline
    $P_\ell + B_0^\mathrm{1loop} + B_2^\mathrm{tree}$ 			&  \makecell{$0.3082$\\$0.311\pm 0.010$}			 & \makecell{$0.6928$\\$0.692\pm 0.011$} 		& 	\makecell{$0.7856$\\$0.794\pm 0.037$} 			& \makecell{$0.1258$\\$0.1255\pm 0.0057$} 			& \makecell{$2.88$\\$2.94\pm 0.11$} 		& \makecell{$0.7962$\\$0.808\pm 0.041$}	\\ \hline
    Planck 							&  $0.3191^{+0.0085}_{-0.016}$ & $0.671^{+0.012}_{-0.0067}$ & $0.807^{+0.018}_{-0.0079}$ 		& $0.1201\pm 0.0013$ 			& $3.046\pm 0.015$ 		& $0.832\pm 0.013$	\\ \hline 

    \end{tabular} \vspace{0.5em}
  \caption{\footnotesize 
   Triangle plots, best-fit values, and relative $68\%$-credible intervals of base cosmological parameters measured from the analysis of BOSS power spectrum multipoles $P_\ell$, $\ell=0,2$, at one-loop, bispectrum monopole $B_0$ at tree or one-loop level, and bispectrum quadrupole $B_2$ at tree-level. 
   Planck $\nu \Lambda$CDM results are shown for comparison. 
   }
  \label{fig:boss}
  \end{figure}

Our main results are summarized in fig.~\ref{fig:boss}, where we plot the posteriors on the cosmological parameters that are effectively scanned. This analysis improves the error bars on the $\Lambda$CDM parameters $\sigma_8$, $h$, and $\Omega_m$ with respect to the power spectrum-only analysis by about 30\%, 18\%, and 13\% respectively, achieving a precision of about $4.7\%$, $1.6\%$, and $3.2\%$ at $68\%$~CL, respectively.\footnote{{Here and in the rest of this work, we quote parameter constraints as the Bayesian $68\%$ credible interval from the one-dimensional marginalized posterior.}}
Notice also that the results improve significantly upon the ones obtained using instead the tree-level prediction for the bispectrum monopole: in particular, $\sigma_8$ is better determined by about $30\%$. 
Naively, a $30\%$ improvement corresponds to doubling the data volume of the survey. 
As it can be seen in the same figure, the results are compatible with the ones obtained with a power-spectrum-only analysis. 
We find no tension with Planck: we measure $\sigma_8$, $h$, and $\Omega_m$ to values consistent at $0.3\sigma, 1.4\sigma, 0.5\sigma$, respectively, with the ones of Planck $\nu\Lambda$CDM~\cite{Planck:2018vyg}.

The paper is organized as follows. In sec.~\ref{sec:data} we describe the data products {and the measurements} we use. In sec.~\ref{sec:theory} we describe the theory model including the observational aspects.  
{In sec.~\ref{sec:likelihood}, we present the likelihood we use to describe the data. }
In sec.~\ref{sec:validation}, we provide some tests for our pipeline. Finally, in sec.~\ref{sec:results}, we provide some additional details about the main results.  Technical aspects {and additional materials} are relegated to the appendices.

\vspace{0.5cm}

\paragraph{A note of warning:} We end this section of the main results with a final note of warning. It should be emphasized that in performing this analysis, as well as the preceding ones using the EFTofLSS by our group~\cite{DAmico:2019fhj,Colas:2019ret,DAmico:2020kxu,DAmico:2020ods,DAmico:2020tty,Zhang:2021yna,DAmico:2022gki}, we have assumed that the observational data are not affected by any unknown systematic error or undetected foregrounds. In other words, we have simply analyzed the publicly available data: the two- and three-point functions of the galaxy density in redshift space as measured from the public galaxy catalogues. Given the additional cosmological information that the theoretical modeling {by} the EFTofLSS allows us to exploit in BOSS data, it might be worthwhile to investigate if potential undetected systematic errors might affect our results. We leave an investigation of these  issues to future work.

%%%%%%%%%%%%%%%%%%%%%%%%%%%%%%%%%%%%%%%%%%%%%%%%%%%%%%%%%%%%%

\section{Data}  \label{sec:data}

\begin{figure}[ht!]
 \centering
  \includegraphics[width=0.99\textwidth]{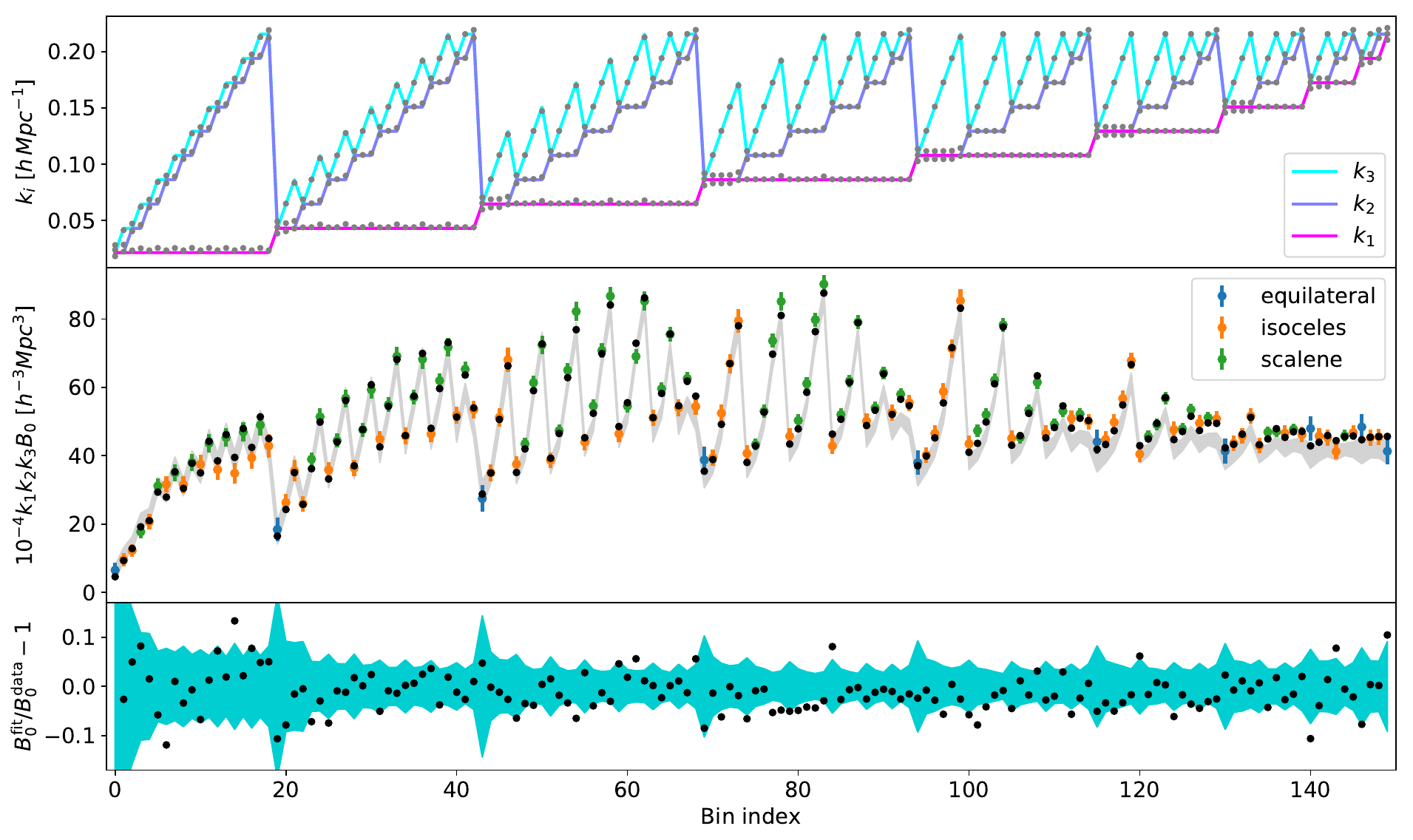}
  \includegraphics[width=0.49\textwidth]{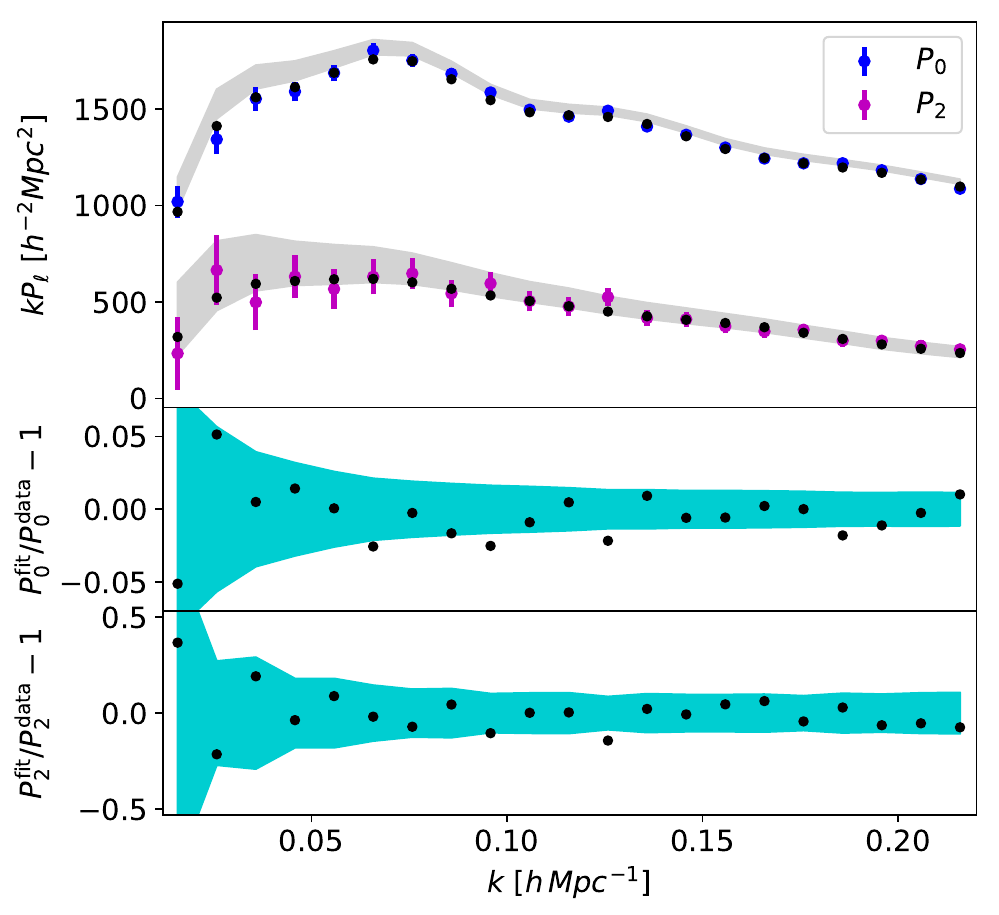}
\includegraphics[width=0.49\textwidth]{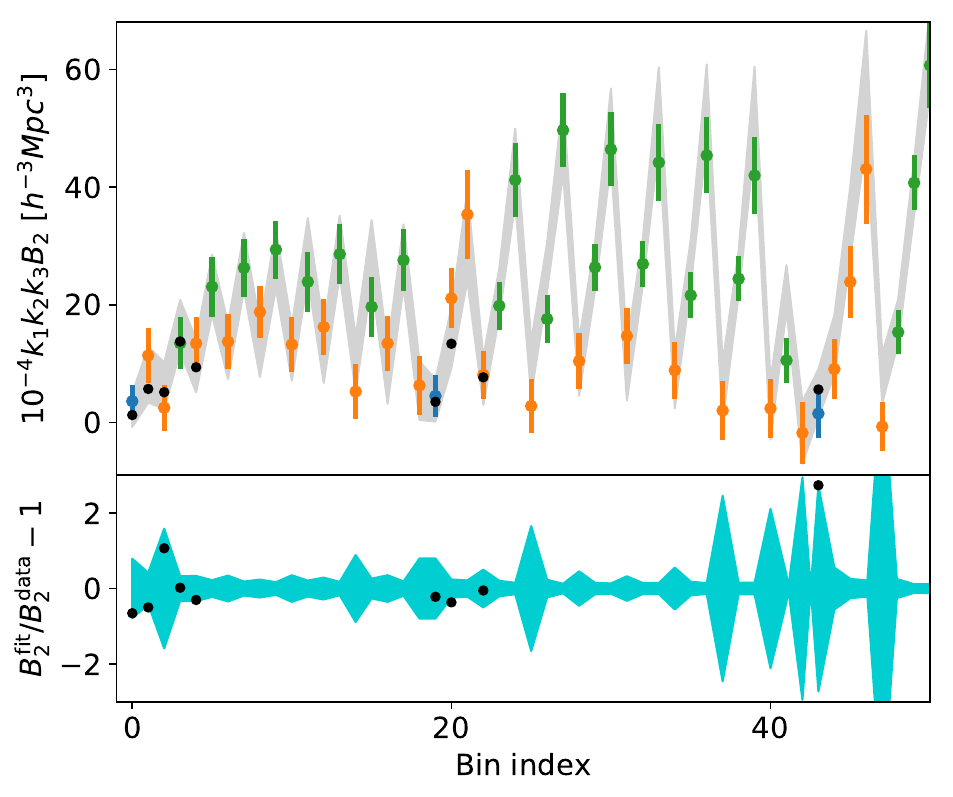}
  \caption{\footnotesize  
   Measurements and best fits of bispectrum monopole $B_0$ (top), power spectrum multipoles $P_\ell$ (bottom left), and bispectrum quadrupole $B_2$ (bottom right) from BOSS (points and error bars) and 2048 Patchy (grey regions) CMASS NGC sky. 
   The bispectrum is shown in bins ordered by their central values forming either an equilateral, isoceles, or scalene triangle, shown in blue, orange, or green, respectively.
   The bin triangle sides (top panel) are shown either by the bin central values (colored lines) or by their effective values (grey points). 
   The best fit (black points) is shown only for the scales analyzed. 
   The relative error bars (turquoise regions) are shown with the best fit residuals for comparison. 
   While only CMASS NGC is shown for clarity, the best fit depicted here is obtained fitting the full combination $P_\ell + B_0 + B_2$ on all BOSS 4 skies. 
   }
  \label{fig:bestfit}
  \end{figure}

\paragraph{BOSS DR12 LRG sample.} 
The main data sample analyzed in this work is the SDSS-III BOSS DR12 luminous red galaxies (LRG) sample~\cite{BOSS:2016wmc}.
We use the BOSS catalogs DR12 (v5) combined CMASS-LOWZ~\cite{Reid:2015gra}.\footnote{Publicly available at \href{https://data.sdss.org/sas/dr12/boss/lss/}{https://data.sdss.org/sas/dr12/boss/lss/}} 
To each galaxy we assign the standard FKP weights for optimality together with the correction weights described in~\cite{Reid:2015gra} for BOSS data and in~\cite{Kitaura:2015uqa} for the patchy mocks. 
The {inverse covariances} are corrected by the Hartlap factor to account for the finite number of mocks used in their estimation~\cite{Hartlap:2006kj}. 
In order to test our analysis pipeline, we will analyze the mean over the 2048 Patchy mocks of CMASS NGC (hereafter referred as `Patchy'). 
We will also make use of the Nseries mocks, which are full $N$-body simulations populated with a Halo Occupation Distribution (HOD) model and selection function similar to the one of BOSS CMASS NGC~\cite{BOSS:2016wmc}.\footnote{Made available at \href{https://www.ub.edu/bispectrum/page11.html}{https://www.ub.edu/bispectrum/page11.html}}
We will analyze the mean of the 84 Nseries realizations (hereafter referred as `Nseries'). 
All celestial coordinates are converted to comoving distance assuming $\Omega_m^{\rm fid} = 0.310$. 

\paragraph{BOSS P+B full-shape measurements.} 
In this work, we analyze the full shape of the power spectrum multipoles $\ell=0,2,$ and of the bispectrum {monopole and quadrupole} (respectively abbreviated `$P_\ell$', `$B_0$' and `$B_2$'). 
Those measurements are shown in fig.~\ref{fig:bestfit} (together with the best fit from our theory model that we discuss later). 
The estimator for the power spectrum is the standard `FKP' estimator~\cite{Feldman:1993ky}, generalized to redshift space in~\cite{Yamamoto:2002bc,Yamamoto:2005dz,Bianchi:2015oia}. 
The bispectrum is estimated using the estimator outlined in~\cite{Gil-Marin:2015sqa} (see also~\cite{Scoccimarro:1997st,Sefusatti:2005br,Baldauf:2014qfa,Scoccimarro:2015bla}). 
The measurements are obtained using the code \code{Rustico}~\cite{Gil-Marin:2015sqa}.\footnote{\href{https://github.com/hectorgil/Rustico}{https://github.com/hectorgil/Rustico}}
For the power spectrum, we find excellent agreement between the measurements from \code{Rustico} and \code{Nbodykit}~\cite{Hand:2017pqn}.\footnote{\href{https://github.com/bccp/nbodykit}{https://github.com/bccp/nbodykit}}
We use \code{Nbodykit} to measure the window functions as described in~\cite{Beutler:2018vpe}, with consistent normalization in the power spectrum as discussed in~\cite{deMattia:2019vdg,deMattia:2020fkb,Beutler:2021eqq}. 

The configurations of the measurements are the following. 
We use a box of side length $L_{\rm box} = 3500\, (2300) \Mpcinvh$ for CMASS (LOWZ), with Piecewise Cubic Spline (PCS) particle assignment scheme and grid interlacing as described in~\cite{Sefusatti:2015aex}. 
The grid is consisting of $512^3$ cells. 
The power spectrum is binned in $\Delta k \simeq 0.01 \hinvMpc$. 
Instead, we bin the bispectrum in $\Delta n = 12\, (9)$ units of the fundamental frequency of the box $k_f$ for CMASS (LOWZ), starting from the bin centered at $n_{\rm min} = 6 + \Delta n/2$, up to the one centered on {$n_{\rm max} = 126\, (69) - \Delta n/2$}, which correspond in frequencies to bins of size $\Delta k = 0.02154 \, (0.02459) \hinvMpc$, with first and last bins centered on $k_{\rm min} = 0.0215 \, (0.029) \hinvMpc$ and {$k_{
\rm max} = 0.215 \, (0.176)  \hinvMpc$}, respectively. 
This choice of bin size is motivated to keep the Hartlap factor at a value safely close to $1$ to limit the effect from the bias of the inverse covariance estimator. 
Given that we have $2048$ patchy mocks at our disposal to estimate the covariance, and, in our analysis, we will analyze $42\, (36)$ $k$-bins in {$P_\ell$} and $150\, (62)$ triangle bins in $B_0$, this makes for the Hartlap factor of about $0.91\, (0.95)$ for CMASS (LOWZ). 
{$B_2$, as analyzed at tree-level, only adds $9$ bins per quadrupole (for both CMASS and LOWZ), which lead to the Hartlap factor of the same order of about $0.9$. }
Importantly, we keep all bins whose centers form a closed triangle.
Explicitly, we {choose the following bins according to their centers ordered as:}
\begin{equation}
  \begin{split}
   (n_1, n_2, n_3)\,, \quad & n_1, n_2, n_3 = n_{\rm min}, n_{\rm min}+dn,  \dots , n_{\rm max}\, , \\
  & \text{if } n_1 \leq n_2 \leq n_3 \text{ and }  n_3 \leq n_1 + n_2 \, .
  \end{split} 
\end{equation}
It follows that there are several bins that contain fundamental triangles that are not closed. How to properly account for them is discussed in sec.~\ref{sec:binning}.

\section{Theory model}  \label{sec:theory} 

Our model for the power spectrum multipoles $P_\ell$, $\ell=0,2$, the bispectrum monopole, $B_0$, and the bispectrum quadrupole, $B_2$, consists in the prediction of EFTofLSS at one loop for $P_\ell$ and $B_0$, and at tree-level for $B_2$. 
We also incorporate a number of observational effects in our modeling to make contact with the measurements.

%%%%%%%%%%%%%%%
%
%

%%%%%%%%%%%%%%%
%
%
\subsection{EFTofLSS at one loop}

In this section, we outline the relevant expressions for the one-loop power spectrum and bispectrum for halos in redshift space $P^{r,h}$ and $B^{r,h}$. The power spectrum and bispectrum are defined as the 2- and 3- point functions of the halo overdensity $\delta_{r,h}$, in Fourier space: 
\bea \label{npdef}
\langle \delta_{r,h} ( \kvec; \hat z ) \delta_{r,h} ( \kvec'  ; \hat z ) \rangle &=& ( 2 \pi)^3 \delta_D ( \kvec + \kvec' )  P^{r,h} ( k ,\hat k \cdot \hat z ) \\ \nonumber
\langle \delta_{r,h} ( \kvec_1 ; \hat z ) \delta_{r,h} ( \kvec_2 ; \hat z )  \delta_{r,h} ( \kvec_3 ; \hat z )\rangle &=& ( 2 \pi)^3 \delta_D ( \kvec_1 + \kvec_2 + \kvec_3 )   B^{r,h} ( k_1 , k_2 , k_3 , \hat k_1 \cdot \hat z , \hat k_2 \cdot \hat z) \ ,
\eea
where $\hat z$ is the line-of-sight direction{, and $\delta_D$ is the Delta dirac function}.  After perturbatively expanding the halo overdensity, we arrive at the following expressions for the one-loop power spectrum:
\be \label{psexpansion}
P^{r,h}_{1\text{-loop tot.}} = P_{11}^{r,h} + (P_{13}^{r,h} +  P_{13}^{r,h,ct} ) + ( P_{22}^{r,h}   + P_{22}^{r,h,\epsilon} )  \ ,
\ee
and the one-loop bispectrum:
\begin{align}
\begin{split} \label{bispexpansion}
B^{r,h}_{1\text{-loop tot.}} & = B_{211}^{r,h} + ( B_{321}^{r,h,(II)} + B_{321}^{r,h,(II),ct}  )  + ( B_{411}^{r,h} + B_{411}^{r,h,ct} )  \\
& \hspace{1in} + ( B^{r,h}_{222} + B_{222}^{r,h,\epsilon} ) + ( B_{321}^{r,h,(I)} +  B_{321}^{r,h, (I), \epsilon} )   \ , 
\end{split}
\end{align}
where we have grouped perturbation theory contributions with the counterterm contributions that renormalize them in parentheses.  
{The loop integrals are evaluated using the techniques described in~\cite{Anastasiou:2022udy}.}
We note that the arguments of the above functions are the same as those given {in eq.~\eqref{npdef}}, and we will often drop the arguments below for clarity.  

Details for all of the above expressions can be found in \appref{theoryapp}, but we summarize the dependence on bias parameters and EFT parameters here for convenience.  For the perturbation theory contributions, we have 
\begin{align}
\begin{split}
& P_{11}^{r,h} [ b_1] \ , \quad  P_{13}^{r,h}   [ b_1 , b_3 , b_8  ]  \ , \quad P_{22}^{r,h} [  b_1 , b_2 , b_5  ]   \ , \\
& B_{211}^{r,h} [  b_1 , b_2 , b_5 ] \ , \quad B_{321}^{r,h,(II)}  [  b_1 , b_2 , b_3, b_5 , b_8  ]  \ , \quad B_{411}^{r,h} [ b_1 , \dots , b_{11} ] \ , \\
&   B^{r,h}_{222}  [ b_1, b_2 , b_5  ] \ , \quad B_{321}^{r,h,(I)} [ b_1 , b_2 , b_3, b_5 , b_6, b_8, b_{10}  ] \ ,  
\end{split}
\end{align}
while for the counterterms, we have
\begin{align}
\begin{split}
& P_{13}^{r,h,ct} [ b_1 , c_{h,1} , c_{\pi,1} , c_{\pi v, 1} , c_{\pi v , 3} ] \ ,   \quad P_{22}^{r,h,\epsilon} [ c^{\rm St}_1 , c^{\rm St}_2 , c^{\rm St}_3]  \ , \\
& B_{321}^{r,h,(II),ct}  [ b_1 , b_2 , b_5 , c_{h,1} , c_{\pi , 1} , c_{\pi v , 1} , c_{\pi v , 3} ]  \ , \quad B_{321}^{r,h,{(I), \epsilon}} [b_1  , c^{\rm St}_1, c^{\rm St}_2 , \{ c^{\rm St}_i \}_{i = 4 , \dots , 13} ]   \ ,  \\
& B_{411}^{r,h,ct} [b_1 , \{c_{h,i} \}_{i=1,\dots,5}  , c_{\pi , 1 } , c_{\pi , 5} , \{ c_{\pi v , j} \}_{j = 1, \dots , 7}   ]  \ , \quad  B_{222}^{r,h,\epsilon}[ c^{(222)}_1, c^{(222)}_2 , c^{(222)}_5 ] \ . 
\end{split}
\end{align}
Notice that the {diagrams} $P_{13}^{r,h} $, $B_{321}^{r,h,(II)} $, and $ B_{411}^{r,h} $ depend on less biases than the kernels in \eqn{kernelbiasdep} would suggest.  This is because, when considering the particular {momentum-}configuration of the kernels that enter the {loop} in \eqn{loopexpressionsrssbias} and \eqn{bispexpressionsrssbias}, they are degenerate with EFT parameters.

{To make contact with our measurements described in~sec~\ref{sec:data}}, what we analyze in the data are various multipoles with respect to the line-of-sight $\hat z$.  In particular, we analyze the power-spectrum {and bispectrum} {monopole and quadrupoles}.  The power-spectrum multipoles are given by 
\be
P_\ell^{r,h} ( k ) = \frac{2 \ell + 1}{2} \int_{-1}^{1} \rmd \mu \, \mathcal{P}_\ell ( \mu ) P^{r,h} ( k , \mu) \ , 
\ee
where $\mathcal{P}_\ell$ are the Legendre polynomials, {and $\mu = \hat k \cdot \hat z$}.
The bispectrum monopole is the average over the {line-of-sight} angles \cite{Scoccimarro:1999ed, Scoccimarro:2015bla, Gil-Marin:2016wya}\footnote{We have corrected a factor of $1/(4 \pi)$ in eq.~(14) of \cite{Gil-Marin:2016wya}.}
\be \label{monopoleeq}
B_0^{r,h} ( k_1 , k_2 , k_3 ) = \frac{1}{4 \pi } \int_{-1}^1 \rmd \mu_1 \int_{0}^{2 \pi} \rmd \phi  \, B^{r,h} ( k_1 , k_2 , k_3, \mu_1 , \mu_2 ( \mu_1 , \phi)  )  \ ,
\ee 
where $\mu_i = \hat k_i \cdot \hat z$, and explicitly, from the triangle conditions:
\begin{align}
\mu_2 ( \mu_1 , \phi)  & = \mu_1 \hat k_1 \cdot \hat k_2 + \sqrt{1 - \mu_1^2} \sqrt{1- (\hat k_1 \cdot \hat k_2)^2} \, \sin \phi \ , \\
\mu_3 ( \mu_1 , \phi ) & = - k_3^{-1} ( k_1 \mu_1 + k_2 \mu_2 ( \mu_1 , \phi)) \ .
\end{align}

The expectation values of the estimator used by \code{Rustico} for the quadrupoles are:  
\begin{align}
\begin{split} \label{bispmultipolesused}
& B_{ {(2,1)}}^{r,h} ( k_1 , k_2 , k_3 ) \equiv  \frac{5}{4 \pi } \int_{-1}^1 \rmd \mu_1 \int_{0}^{2 \pi} \rmd \phi   \,  \mathcal{P}_2 ( \mu_1 ) B^{r,h} ( k_1 , k_2 , k_3, \mu_1 , \mu_2 ( \mu_1 , \phi)   )   \ , \\
& B_{{(2,2)}}^{r,h} ( k_1 , k_2 , k_3 ) \equiv  \frac{5}{4 \pi } \int_{-1}^1 \rmd \mu_1 \int_{0}^{2 \pi} \rmd \phi   \,  \mathcal{P}_2 ( \mu_2 ( \mu_1 , \phi)  ) B^{r,h} ( k_1 , k_2 , k_3, \mu_1 , \mu_2 ( \mu_1 , \phi)   )   \ , \\
& B_{{(2,3)}}^{r,h} ( k_1 , k_2 , k_3 ) \equiv  \frac{5}{4 \pi } \int_{-1}^1 \rmd \mu_1 \int_{0}^{2 \pi} \rmd \phi   \,  \mathcal{P}_2 ( \mu_3 ( \mu_1 , \phi)  ) B^{r,h} ( k_1 , k_2 , k_3, \mu_1 , \mu_2  ( \mu_1 , \phi)  )   \ .
\end{split}
\end{align}
We work directly in this basis of quadrupoles, that are linear combinations of the $B_{2m}$ coefficients of the spherical-harmonics expansion defined in~\cite{Scoccimarro:1999ed}. 
{We note that if only considering the bispectrum monopole,  $c_2^{(222)}$  and  $c_5^{(222)}$ become degenerate, so we redefine $c_2^{(222)} \rightarrow c_2^{(222)}  - c_5^{(222)} /6$~(\footnote{{When considering} in addition the bispectrum quadrupole at one loop, this degeneracy breaks. }).}

%%%%%%%%%%%%%%%%%%%%%%
%
%

\subsection{IR-resummation}

The IR-resummation is a crucial effect to include in our theory model, {in order to correctly reproduce the BAO}.
For the power spectrum, we use the full resummation of~\cite{Senatore:2014via} as implemented in Pybird~\cite{DAmico:2020kxu}.
For the bispectrum instead we rely on a wiggle-no wiggle approximation, following~\cite{Ivanov:2018gjr}.
For the linear bispectrum, the formula we implement is: 
\begin{equation}
B_{211}^{r,h} = 2  K_1^{r,h} ( \kvec_1 ; \hat z )  K_1^{r,h} ( \kvec_2  ; \hat z )  K_2^{r,h} ( \kvec_1, \kvec_2 ;  \hat z ) P_{\rm LO} ( k_1 ) P_{\rm LO} ( k_2 ) + \text{ 2 perms.} \ ,
\end{equation}
where
\begin{equation}
  P_{\rm LO}(k) = P_{\rm nw}(k) + (1 + k^2 \Sigma_{\rm tot}^2) e^{- {k^2} \Sigma_{\rm tot}^2} P_{\rm w}(k) \, .
\end{equation}
Here $P_{\rm w}(k) = P_{11}(k) - P_{\rm nw}(k)$, and $P_{\rm nw}(k)$ is the no-wiggle power spectrum, which we obtain using the sine-transform algorithm described in~\cite{2010JCAP...07..022H} and detailed in~\cite{Chudaykin:2020aoj}. 
Then $\Sigma_{\rm tot}^2$ is defined by 
\begin{align}
  &\Sigma^{{2}}_{\rm tot} = - \frac{2}{15} f^2 \, \delta \Sigma^2 + \(1 + \frac{1}{3} f (2 + f) \) \Sigma^2 \, ,\\
  &\Sigma^2 = \frac{4 \pi}{3} \int_0^{\Lambda} {\frac{\rmd q}{(2 \pi)^3}} P_{\rm nw}(q) \left[ 1 - j_0(q x_{\rm osc}) + 2 j_2(q x_{\rm osc}) \right] \, , \\
  &\delta \Sigma^2 = 4 \pi \int_0^{\Lambda} {\frac{\rmd q}{(2 \pi)^3}} P_{\rm nw}(q) j_2(q x_{\rm osc}) \, ,
\end{align}
where we choose $\Lambda = 1 \, \hinvMpc$ and $x_{\rm osc} = 110 \, \Mpcinvh$, and $j_l$ are the spherical Bessel functions.\footnote{{We have checked that changing $\Lambda$ to $0.12 \hinvMpc$ leads to insignificant differences in the posteriors. }}
For the loop, our choice is to only substitute the non-integrated $P_{11}(k)$ with
\begin{equation}
  P_{\rm NLO}(k) = P_{\rm nw}(k) + e^{- {k^2} \Sigma_{\rm tot}^2} P_{\rm w}(k) \, ,
\end{equation}
while {for linear power spectra whose argument are being integrated, }we use $P_{11}$. We discuss the goodness of this approximation in sec.~\ref{sec:validation}.  {Another method of BAO damping for the tree-level bispectrum was given and tested on Patchy mocks in \cite{Sugiyama:2020uil}.} 

\subsection{Window function}

The power spectrum and bispectrum need to be convolved with the window function of the survey.
For the power spectrum, this is standard and does not present numerical challenges.
However, for the bispectrum this becomes more challenging.
Therefore, we resort to an approximation used in~\cite{Gil-Marin:2014sta}, which amounts to evaluating the linear bispectrum with the windowed power spectrum.
In formula, we have:
\begin{equation}\label{eq:window}
  B_{211}^{r,h} = 2  K_1^{r,h} ( \kvec_1 ; \hat z )  K_1^{r,h} ( \kvec_2  ; \hat z )  K_2^{r,h} ( \kvec_1, \kvec_2 ;  \hat z ) [W \ast P_{11}](\vec k_1) [W \ast P_{11}]( \vec k_2 ) + \text{ 2 perms.} \ ,
\end{equation}
where $[W \ast P_{11}](\vec k) = \int \frac{\rmd^3 k'}{(2 \pi)^3} W(\kvec - \kvec') P_{11}(\vec k')$. Rather than projecting (\ref{eq:window}) into multipoles, we project (\ref{eq:window}) as if there was no window function, and for $[W \ast P_{11}](\vec k)$ we use the following: for the monopole, we use $W\rightarrow W_{00}$ and, for the quadrupole we use $W\rightarrow W_{22}$, where $W_{00}$ and $W_{22}$ are defined in~\cite{DAmico:2019fhj}. Because the window is a small effect, we do not apply it to the loop bispectrum.
 We discuss the goodness of this approximation in sec.~\ref{sec:validation}.

\subsection{Alcock-Paczynski effect}

To estimate the galaxy spectra from data, a reference cosmology is assumed to transform the measured redshifts and celestial coordinates into three-dimensional cartesian coordinates.
The difference between the reference cosmology and the true cosmology produces a geometrical distortion known as the Alcock-Paczynski (AP) effect~\cite{Alcock:1979mp}.
We introduce the transverse and parallel distortion parameters:
\begin{equation}
  q_{\perp} = \frac{D_A(z) H_0}{D_A^{\rm ref}(z) H_0^{\rm ref}} \, , \qquad
  q_{\parallel} = \frac{H^{\rm ref}(z) / H_0^{\rm ref}}{H(z) / H_0} \, ,
\end{equation}
where $D_A$ is the angular diameter distance, and the factors of $H_0$ are there since our wavenumbers are in units $\hinvMpc$.
In terms of these, the true wavenumber and angle with the line of sight are related to the ones in the reference cosmology by:
\begin{align}
  k = \frac{k^{\rm ref}}{q_\perp} \left[ 1 + (\mu^{\rm ref})^2 \( \frac{1}{F^2} - 1 \) \right]^{1/2} \, , \quad
  \mu = \frac{\mu^{\rm ref}}{F} \left[ 1 + (\mu^{\rm ref})^2 \( \frac{1}{F^2} - 1 \) \right]^{-1/2} \, ,
\end{align}
where $F = q_{\parallel} / q_\perp$.
To match the measured power spectrum multipoles, we do the following integral: 
\begin{equation}
  P_{\ell}(k^{\rm ref}) = \frac{2 \ell + 1}{2 q_{\parallel} q_{\perp}^2} \int_{-1}^1 \rmd \mu^{\rm ref} \mathcal{L}_{\ell}(\mu^{\rm ref}) P(k( k^{\rm ref}, \mu^{\rm ref}), \mu(\mu^{\rm ref})) \, .
\end{equation}
{The formula for the bispectrum is:
\begin{equation}\label{eq:AP}
   B_{(\ell,i)} (k_1^{\rm ref}, k_2^{\rm ref}, k_3^{\rm ref}) = \frac{{2\ell+1}}{2 q_{\parallel}^2 q_{\perp}^4} \int_{-1}^1 \rmd \mu_1^{\rm ref} \int_{0}^{2 \pi} \frac{\rmd \phi^{\rm ref}}{2 \pi} B(k_1, k_2, k_3, \mu_1, \mu_2, \mu_3) \mathcal{P}_\ell(\mu_i) \, .
\end{equation} }

{For the bispectrum, we only apply the Alcock-Paczynski effect on the tree-level part, as it is a small effect: we find that, within BOSS error bars (on $\Omega_m$), it is an effect of at most $\sim 1\sigma$, and accordingly, the change in $\chi^2$ is at most $1$ if neglecting it completely. Given the size of the loop terms and counterterms, it is thus safe to neglect it there. 
We find that we can achieve sufficient numerical accuracy using a nested trapezoidal rule with only $13$ points in $\mu$ and $4$ points in $\phi$, after using symmetries to restrict the integration domain to $\mu_1 \in [0, 1]$, and $\phi \in [-\pi/2, \pi/2]$. 

\subsection{Binning} \label{sec:binning}

For the power spectrum, data are an average over {spherical} shells in momentum space.
The theoretical prediction needs therefore to be averaged over the fundamental modes of the chosen grid.
Since our bins have many fundamental modes, in practice we do an integral of the power spectrum over a bin, which is numerically very simple.
The effect of binning is anyway small for the power spectrum, with respect to the error bars of our data and simulations.

For the bispectrum, we have an average over fundamental (closed) triangles in a bin of width $\Delta k$ around a central triangle with sides $k_1$, $k_2$, $k_3$.
Especially for our chosen bins with $\Delta k \simeq 0.02$, it is important to take into account the binning effects when comparing the theory to the data.
The average should be done as a sum over fundamental triangles:
\begin{equation}
B^{r,h}_{ {(\ell,i)}, \rm bin}(k_1, k_2, k_3) = \frac{2 \ell +1}{N_T} \sum_{\qvec_1 \in k_1} \sum_{\qvec_2 \in k_2} \sum_{\qvec_2 \in k_2} \delta_K(\qvec_1+\qvec_2+\qvec_3) B^{r,h}(\qvec_1, \qvec_2, \qvec_3) \mathcal{P}_\ell(\mu_i)\, ,
\end{equation}
{and we note that, here and elsewhere, $B^{r,h}(\qvec_1, \qvec_2, \qvec_3)$ is the full redshift-space bispectrum, i.e. we have suppressed the dependence on $\hat z$ for notational convenience.}
{Here} $N_T$ is the number of fundamental triangles in the bin, $\delta_K$ is the Kronecker delta function, and the notation $\qvec_i \in k_i$ means a sum over the fundamental modes $\qvec_i$ for which $k_i - \frac{\Delta k}{2} \leq |\qvec_i| < k_i + \frac{\Delta k}{2}$.
Calculating such a sum is numerically very challenging.
However, since in each bin there are many fundamental triangles, we expect that an integral approximation should work well.
The only caveat is that one needs to integrate only over the closed triangles.
In particular, this is very important for bins such that {$k_3 + \Delta k /2 > k_1 + k_2 - \Delta k$ (remember that our ordering is $k_1 \leq k_2 \leq k_3$)}, for which there are {configurations of modes that do not form a closed triangle} in the bin. 

Therefore, we implement the following formula:
\begin{align}
\begin{split}
  B^{r,h}_{ {(\ell,i)} , \rm bin}(k_1, k_2, k_3) & = \frac{{2\ell+1}}{V_T} \left( \prod_{i=1}^3 \int_{V_i} \frac{\rmd^3 q_i}{(2 \pi)^3}  \right) (2 \pi)^3 \delta_D^{(3)}(\qvec_1+\qvec_2+\qvec_3)  B^{r,h}(\qvec_1, \qvec_2, \qvec_3) \mathcal{P_\ell}(\mu_i) \, ,
\end{split}
\end{align}
where
\begin{equation}
  V_T \equiv \left( \prod_{i=1}^3 \int_{V_i} \frac{\rmd^3 q_i}{(2 \pi)^3}  \right)  (2 \pi)^3 \delta_D^{(3)}(\qvec_1+\qvec_2+\qvec_3) \, ,
\end{equation}
and we used the notation 
\begin{equation}
  \int_{V_i} \frac{\rmd^3 q_i}{(2 \pi)^3} = \int_{k_i } \frac{\rmd q_i}{2 \pi^2} q_i^2 \int \frac{\rmd^2 \hat{q}_i}{4 \pi} \  , \quad \text{where} \quad \int_{k_i }   \equiv \int_{k_i - \frac{\Delta k}{2}}^{k_i + \frac{\Delta k}{2}} \rmd q_i  \ .
\end{equation}
As shown in app.~\ref{app:binningd}, we can perform the angular integrals and find  
\begin{equation} \label{eq:binning} 
  B^{r,h}_{{(\ell,i)}, \rm bin}(k_1, k_2, k_3) = \frac{1}{V_T} \int_{k_1 } \rmd q_1 \int_{k_2}  \rmd q_2 \int_{k_3  } \rmd q_3 \,   q_1 q_2 q_3 \,  \frac{\beta\(\Delta_q\)}{8 \pi^4} B^{r,h}_{{(\ell,i)}}(q_1, q_2, q_3) \,  \, ,
\end{equation}
\begin{equation}
  V_T =\int_{k_1 } \rmd q_1 \int_{k_2}  \rmd q_2 \int_{k_3  } \rmd q_3  \, q_1 q_2 q_3 \, \frac{\beta\(\Delta_q\)}{8 \pi^4}  \, ,
\end{equation}
where {$\beta(\Delta_q) = 1/2$ if $q_1$, $q_2$, $q_3$ form a folded triangle, $\beta(\Delta_q) = 1$ for all other (closed) triangles, and $\beta(\Delta_q) = 0$ otherwise. }

{We apply only the binning in this way to the tree-level part. 
For efficient numerical evaluation of the integrals in eq.~\eqref{eq:binning}, we implement the bispectrum binning as follow. 
For a given bin centered in $(k_1, k_2, k_3)$, we enforce that $(q_1, q_2, q_3)$ forms a triangle by redefining the integration boundaries: $q_1 \in [k_1-\Delta k/2, k_1 + \Delta k/2]$, $q_2 \in [k_2-\Delta k/2, k_2 + \Delta k/2]$, and $q_3 \in [|k_1-k_2|, k_1+k_2]$. 
Whenever $q_3$ can not satisfy this triangle inequality, we drop this configuration. 
As such, we can drop the $\beta(\Delta_q)$ function inside the integral. 
We perform a change of variable $q_3 \rightarrow \cos(\theta_{12}) \equiv (q_3^2 - q_1^2 - q_2^2) / (2 q_1 q_2)$, such that the integral measure becomes $q_1 q_2 q_3 \ \rmd q_1 \rmd q_2 \rmd q_3 \rightarrow q_1^2 q_2^2 \ \rmd q_1 \rmd q_2  \rmd \cos(\theta_{12})$, 
We then discretize the integration domain in 6 evenly-spaced points in $q_1$, 6 in $q_2$, and 4 in $\cos(\theta_{12})$. 
On each point of this grid, we evaluate the $14$ pieces of the tree-level part of the bispectrum. 
The binning integrals are then performed with a nested trapezoidal rule over the grid. 
Given that the AP integrals, eq.~\eqref{eq:AP}, need to be performed for each of those evaluations, we limit the number of evaluations by first looking for (and storing) the common triangles of the discretized domains over all the bins we need to evaluate. 
For our $150$ bins in CMASS, this reduces the total number of evaluations by about a factor $1.5$, from $6 \cdot 6 \cdot 4 \cdot 150 = 21600$ to $13782$. 
After compilation of the integrand expressions going in the AP integrals, we are able to evaluate the binned bispectrum in our Python code with an overall runtime of $\sim 1$ second on $1$ CPU. 
The numerical precision of such evaluation has been extensively tested, in particular against Monte-Carlo integrations, and is found to be under control for the data and simulations error bars we analyze in this work.   
}

The loop pieces and counterterms, that are small with respect to the linear term, are {instead} evaluated on effective wavenumbers.\footnote{We checked that binning the loop did not lead to appreciable different posteriors.}
They are defined, as described in~\cite{Rizzo:2022lmh}, by the following averages: 
\begin{align}
  & k_{\rm eff, 1} = \frac{1}{V_T} \int_{k_1} \frac{\rmd q_1}{2 \pi} \int_{k_2} \frac{\rmd q_2}{2 \pi} \int_{k_3} \frac{\rmd q_3}{2 \pi} \, q_1 q_2 q_3 \, \beta\(\Delta_q\) \min(q_1, q_2, q_3) \, , \\
  & k_{\rm eff, 2} = \frac{1}{V_T} \int_{k_1} \frac{\rmd q_1}{2 \pi} \int_{k_2} \frac{\rmd q_2}{2 \pi} \int_{k_3} \frac{\rmd q_3}{2 \pi}  \, q_1 q_2 q_3 \, \beta\(\Delta_q\) \mathrm{med}(q_1, q_2, q_3) \, , \\
  & k_{\rm eff, 3} =\frac{1}{V_T} \int_{k_1} \frac{\rmd q_1}{2 \pi} \int_{k_2} \frac{\rmd q_2}{2 \pi} \int_{k_3} \frac{\rmd q_3}{2 \pi}  \, q_1 q_2 q_3 \, \beta\(\Delta_q\) \max(q_1, q_2, q_3) \, .
\end{align}
{As expected from the size of those terms and the size of the binning effect ($\sim 1\sigma$), we have checked that properly binning the loop instead of evaluating them on these effective wavenumbers lead to negligible shift in the $\min \chi^2$ and in the posteriors for the analyses presented in this work.}

%%%%%%%%%%%%%%%%%%%%%%%%%%%%%%%%%%%%%%%%%%%%%%%%%%%%%%%%%%%%%
\section{Likelihood}\label{sec:likelihood}

To analyze the data, we start from a Gaussian likelihood, which is multiplied by the prior to arrive at the Bayesian posterior $\calP$ over cosmological and bias parameters:
\begin{equation}
  - 2 \ln \calP = (T_i - D_i) C^{-1}_{ij} (T_j - D_j) - 2 \ln \calP_{\rm pr} \, ,
\end{equation}
where $T_i$ is the full vector of theory predictions in bin $i$, containing power spectrum multipoles and bispectra, $D_i$ the corresponding data measurement in bin $i$, $C_{ij}$ is the full covariance between bins $i$ and $j$, and $\calP_{\rm pr}$ is a generic prior on the parameters.

Our theory model depends on cosmological and EFT parameters.
It is the case that many EFT parameters appear linearly in the theory model.
Denoting them by $g_\al$, we can write
\begin{equation}
  T_i = g_\al T_{G, i}^\al + T_{NG, i} \ ,
\end{equation}
where $T_{G, i}^\al$ and $T_{NG, i}$ depend non-linearly (that is, at least quadratically) on the other cosmological parameters and {three} biases {for each sky cut}.
Since we are interested in the marginalized posteriors over cosmological parameters, it is very convenient to do the analytical Gaussian integration over the $g_{\al}$.
We will also choose a Gaussian prior on them, with covariance $\sigma_{\al \bt}$ and mean $\hat{g}_\al$.\footnote{We only use $\hat{g}_\al \neq 0$ in one of the checks in sec.~\ref{sec:validation}.}
Collecting the powers of $g_{\al}$, we can write the posterior in the following form:
\begin{equation}
  - 2 \ln \calP = g_\al F_{2, \al \bt} g_\bt - 2 g_\al F_{1, \al} + F_0 \, ,
\end{equation}
where the $F$'s are defined as:
\begin{align}
  & F_{2, \al \bt} = T_{G, i}^\al C^{-1}_{ij} T_{G, j}^\bt + \sigma^{-1}_{\al \bt} \, , \label{eq:F2def} \\
  & F_{1, \al} = - T_{G, i}^\al C^{-1}_{ij} (T_{NG, j} - D_j) + \sigma^{-1}_{\al \bt} \hat{g}_\bt \, , \\
  & F_{0} = (T_{NG, i} - D_i) C^{-1}_{ij} (T_{NG, j} - D_j) + \hat{g}_\al \sigma^{-1}_{\al \bt} \hat{g}_\bt - 2 \ln \Pi \, ,
\end{align}
where $\Pi$ is a generic prior on the cosmological and bias parameters non analytically marginalized. In other words, we assume that $\calP_{\rm pr}$ is a sum of a Gaussian prior over the $g_\alpha$ and a remaining prior on the other parameters.
After integrating the $g_\al$, we have the marginalized posterior:
\begin{equation}
  - 2 \ln \calP_{\rm marg} = - F_{1, \al} F_{2, \al \bt}^{-1} F_{1, \bt} + F_0 + \ln \det \(\frac{F_2}{2 \pi} \) \, .
\end{equation}

\paragraph{Prior.} 
In our analysis, we vary the cosmological parameters $\omega_{cdm}$, $h$, and $\ln \(10^{10} A_s\)$ with a flat uninformative prior, while we use a Gaussian prior on $\omega_b$ of mean $\omega_{b, \textrm{BBN}} = 0.02233$ and standard deviation $\sigma_{\rm BBN} = 0.00036$, motivated from Big-Bang Nucleosynthesis (BBN) experiments~\cite{Mossa:2020gjc}.
We instead fix $n_s$ to the truth of the simulations or to the Planck preferred value when analyzing the data~\cite{Planck:2018vyg}. 
When analyzing the BOSS data, we also fix the neutrino to minimal mass following Planck prescription.\footnote{As we describe later in sec.~\ref{sec:validation}, we add a linear prior on $\Omega_m$, $h$ and $b_1$ in order to mitigate phase-space projection effects.}

The EFT parameters should instead be restricted to be ${\cal{O}}(1)$ numbers, for consistency of the perturbative expansion. 
The EFTofLSS is an expansion in the size of fluctuations and derivatives.
Both of these are suppressed by a nonlinear scale $k_{\rm NL} \simeq k_{\rm M} \simeq 0.7 \hinvMpc$, where $k_{\rm NL}$ is the nonlinear scale for the matter field, and $k_{\rm M}$ is the typical wavenumber associated to galaxy size.
However, it was recognized in~\cite{Lewandowski:2015ziq,DAmico:2021ymi} that terms involving expectation values of velocity fields, coming from the transformation to redshift space, define a new scale, which we denote by $k_{\rm NL,R} \simeq k_{\rm NL} / \sqrt{8}$.
We therefore write down each operator in the EFT expansion with either a $k_{\rm M}$ or a $k_{\rm NL,R}$ suppression, depending on its origin.
We then use a Gaussian prior of width 2 centered on 0 on all the EFT parameters that we analytically marginalize, with the following exception:
on $c_{h,1}, c_{\pi,1}, c_{\pi v,1}$, and $c_2^{\rm St}$, that already appear in the power spectrum (see their definitions in app.~\ref{theoryapp}), we put instead a Gaussian prior of width 4 centered on 0, such that the prior is the same as the ones used in our previous series of analyses with the power spectrum only (see e.g.~\cite{DAmico:2019fhj,Colas:2019ret,Zhang:2021yna}). 
For the quadratic biases, we define the linear combinations $c_2 = (b_2 + b_5) / \sqrt{2}$, {$c_4 = (b_2 - b_5) / \sqrt{2}$}, and we assign on them a Gaussian prior of width 2 centered on 0.
{Finally, for} $b_1$, which is positive definite, we use a lognormal prior of mean $0.8$ (since $e^{0.8} = 2.23$), and variance 0.8, such that $[0, 3.4]$ is the $68\%$ bound for this prior on $b_1$. {For definiteness, in our prior, we take $k_{\rm NL} = k_{\rm M} = 0.7 \hinvMpc$ and $\bar n=4\cdot 10^{-4} ({\rm Mpc}/h)^3$}.

When analyzing more than one sky, we can use the information that the bias and EFT parameters should be the same at the same redshift, and their time evolution is expected to be {comparable to the growth factor to some small power. 
This allows us to estimate the variation of $b_1$ between CMASS or LOWZ effective redshifts, to be about 20\%. }
Therefore, in our multisky analyses the biases are correlated, which, as explained in the following section, helps to mitigate prior volume effects.
In practice, let us consider the 4-sky analysis and the $b_1$ parameters, which will be a vector {$(b_1^{(1)}, b_1^{(2)}, b_1^{(3)}, b_1^{(4)})$, with one $b_1^{(i)}$ for each sky.}
The prior on it is a multivariate lognormal with correlation matrix:
\begin{equation}
  \begin{pmatrix}
    1 & \rho_{12} & \rho_{13} & \rho_{12} \rho_{13} \\
    \rho_{12} & 1 & \rho_{12} \rho_{13} & \rho_{13} \\
    \rho_{13} & \rho_{12} \rho_{13} & 1 & \rho_{12} \\
    \rho_{12} \rho_{13} & \rho_{13} & \rho_{12} & 1 
  \end{pmatrix} \, ,
\end{equation}
where $\rho_{ij} = 1 - \eps_{ij}^2/2$, and we choose $\eps_{12} = 0.1$, $\eps_{13} = 0.2$. 
This formula is motivated by the fact that two variables distributed according to a bivariate normal with correlation $\rho$, the standard deviation of the difference is $\eps = \sqrt{2 (1- \rho)}$. Our choices of $\eps_{ij}$ then reflect that we expect the values of $b_1$ to be different only by about $10\%$ between NGC and SGC, given slightly different selection function, and only by about $20\%$ between CMASS and LOWZ, given the redshift evolution of $b_1$. We use the same correlation matrix for {the Gaussian priors on} all the quadruplets $c_2$, $c_4$ and the $g_\al$ parameters.

\paragraph{Posterior sampling.} 

Our analyses are performed using the Metropolis-Hastings sampler as implemented in \code{MontePython 3}~\cite{Brinckmann:2018cvx}, with the theory model evaluated using \code{CLASS}~\cite{Blas:2011rf} and \code{PyBird}.
We declare our MCMC converged when the Gelman-Rubin criterion~\cite{Gelman:1992zz} is $\leq 0.02$.
The plots and summary statistics are calculated with the \code{GetDist}~\cite{Lewis:2019xzd} package.

\section{Pipeline validation\label{sec:validation}}

For our analyses, we use the following scale cut: 
$k_{\rm min} = 0.01 \hinvMpc$ for all observables, $\kmax= 0.23 \hinvMpc$ for $P_\ell$ and $B_0$, and $\kmax = 0.08 \hinvMpc$ for $B_2$, on CMASS. 
For LOWZ instead, we use  $\kmax= 0.20 \hinvMpc$ for $P_\ell$ and $B_0$ }, following~\cite{DAmico:2019fhj}. We keep $\kmax = 0.08 \hinvMpc$ for $B_2$ on LOWZ.\footnote{{For comparison purpose, we sometimes fit $B_0$ using the tree-level prediction instead. When doing so, $B_0$ is denoted $B_0^{\rm tree}$ (to distinguish from $B_0^{\rm 1loop}$) and is fitted up to $\kmax = 0.08 \hinvMpc$ for both CMASS and LOWZ. }}
In this section, we perform multiple checks to validate our method at this scale cut.  

\subsection{Measuring and fixing phase-space effects} 

\begin{figure}[t!]
  \centering
  \includegraphics[width=0.49\textwidth]{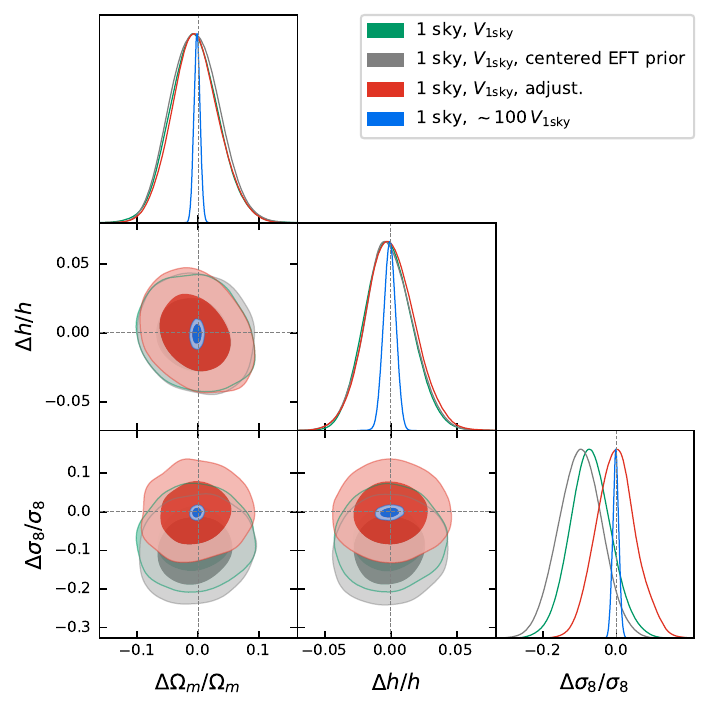}
  \includegraphics[width=0.49\textwidth]{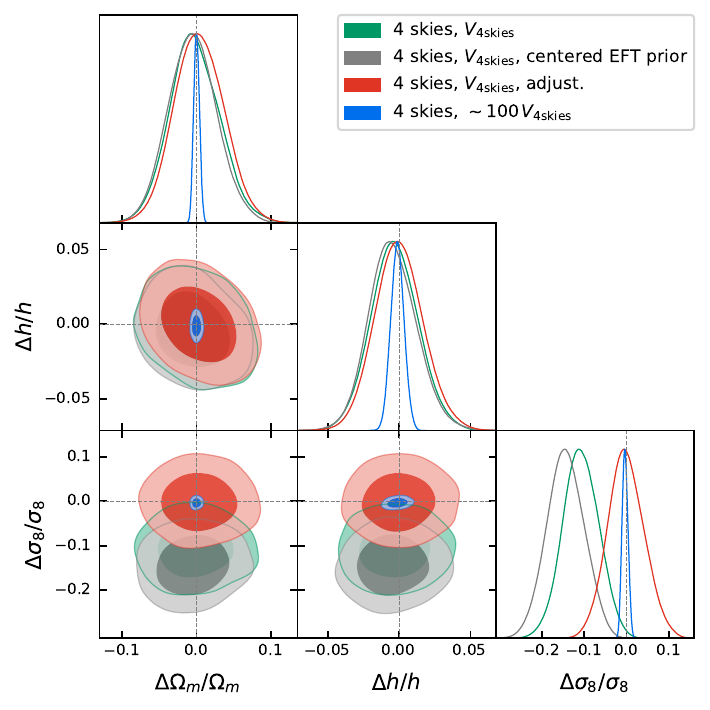}
    \vspace{0.5em}
    \scriptsize
    
\begin{tabular}{|c|c|c|c|c|c|c|} \hline
 $\sigma_{\rm proj}/\sigma_{\rm stat}$       & $\Omega_m$ & $h$  & $\sigma_8$ & $\omega_{cdm}$ & $\ln(10^{10}A_s)$  & $S_8$ \\ \hline
1 sky, $\sim 100\, V_{\rm 1sky}$ &  -0.1 & -0.14 & -0.21  & -0.2 & -0.07 & -0.23  \\ 
1 sky, $V_{\rm 1sky}$, adjust. & 0.13 & 0.06 & 0.04 & 0.15 & -0.04 & 0.08           \\ 
4 skies, $V_{\rm 4skies}$, adjust.  & 0.1 &  0.  & -0.05 &  0.07 &-0.06 &-0.01          \\ \hline
\end{tabular}

  \caption{\footnotesize 
 Triangle plots of base cosmological parameters obtained fitting synthetic data analyzed using a covariance with BOSS volume $V_{\rm BOSS}$ or rescaled to a large volume $\sim 100 V_{\rm BOSS}$, with prior on the EFT parameters centered on their truth, or with phase-space projection adjustment. 
  Here the synthetic data are corresponding exactly to our model $P_\ell + B_0 + B_2$ on the best fit of patchy. 
 `1 sky'  or `4 skies' correspond to CMASS NGC or all BOSS skycuts, respectively. 
 The grey lines in the triangle plots represent the truth. 
  We also show the relative deviations $\sigma_{\rm proj}/\sigma_{\rm stat}$ on the base cosmological parameters from the truth from those various analyses.  
  {In summary, the addition of a phase-space correction prior to our likelihood allows us to recover unbiased mean in the 1D posteriors of the cosmological parameters of interest. }  }
  \label{fig:synth}
  \end{figure}
  
 \begin{table}[t!]
\scriptsize
\centering
\begin{tabular}{|l|l|c|c|c|c|c|c|} \hline
  \multicolumn{2}{|c|}{$\sigma_{\rm proj}/\sigma_{\rm stat}^{\rm data}$}       & $\Omega_m$ & $h$  & $\sigma_8$ & $\omega_{cdm}$ & $\ln(10^{10}A_s)$ & $S_8$ \\ \hline
 \multirow{3}{*}{\begin{turn}{90}Nseries\end{turn}}  
 & $P_\ell$ & -0.02 &  0.05 & 0.08 & 0.02 & 0.05 & 0.07 \\
 & $P_\ell + B_0$ & -0.06 & -0.03 & -0.04 & -0.08 &  0.03 &-0.06 \\
 & $P_\ell + B_0+B_2$ & -0.12 &-0.  & -0.04 &-0.11 &  0.04 & -0.08 \\ \hline
  \multirow{3}{*}{\begin{turn}{90}1 sky\end{turn}}  
 & $P_\ell$ & -0.15 & 0.07 & -0.11 & -0.06 & -0.08 & -0.15 \\
 & $P_\ell + B_0$ & 0.07 & 0.06 & 0.09 & 0.11 & 0.02 & 0.1 \\
 & $P_\ell + B_0+B_2$ & 0.13 & 0.06 & 0.04 & 0.15 & -0.04 & 0.08 \\ \hline
   \multirow{3}{*}{\begin{turn}{90}4 skies\end{turn}} 
 & $P_\ell$ & -0.01 & 0.05 &-0.03 & 0.02 &-0.04 &-0.03 \\
 & $P_\ell + B_0$ & 0.05 & -0.  &  0.01 & 0.03 & 0.01 & 0.03 \\
 & $P_\ell + B_0+B_2$ & 0.1 &  0.  & -0.05 &  0.07 &-0.06 &-0.01 \\  \hline
\end{tabular}
\caption{\footnotesize Residual deviations $\sigma_{\rm proj}$ after phase-space projection adjustment measured on synthetic data generated and fitted with our model $P_\ell+B_0+B_2$ with truth given by the best fits of Nseries, Patchy 1 sky, or Patchy 4 skies, relative to BOSS error bars $\sigma_{\rm stat}^{\rm data}$. }
\label{table:synth}
\end{table}

Our likelihood has several EFT parameters on top of the cosmological parameters. Some of these appear in the likelihood in a Gaussian way, and we analytically marginalize over them. Performing such a Gaussian integral corresponds to putting these parameters to their best fit values, given all the other parameters and observational data. At this point, we are left with a likelihood which has a non-Gaussian dependence on the EFT and the {cosmological} parameters. 
  
Now, there is an interesting phenomenon that we would like to describe. Let us analyze data that are generated with our theory model: {the EFTofLSS plus observational effects as described in sec~\ref{sec:theory}}. We refer to these as {`synthetic'} data. We generate these synthetic data by choosing the best fit EFT parameters that we find by fitting the average of $2048$ Patchy simulations, on the Patchy cosmology, so that the resulting EFT and {cosmological} parameters are at realistic values. In this case, the best fit has $\chi^2=0$ once we put flat priors on the EFT parameters, and we should clearly recover the correct cosmological parameters. However, as it can be seen from fig.~\ref{fig:synth}, in green, the {sampled} posteriors show biases in all {1D posteriors of the cosmological} parameters, and in particular in $\sigma_8$ and $\Omega_m$. What is going on?

The first hypothesis is that there could be an error in our pipeline. This hypothesis can be discarded by noticing that if we analyze the data with a covariance that is about 100 times smaller, we recover the truth with exquisite precision (see the blue curve in fig.~\ref{fig:synth}). So, we exclude this hypothesis.

Another reason for the offset of the {green} curve in fig.~\ref{fig:synth} could be the prior on the EFT parameters. 
In fact, while on the synthetic data the EFT parameters have some definite values (which are well within the priors), our Gaussian priors are centered at zero, and so the true value of the EFT parameters are slightly disfavored by the priors. 
{We check if this can be the reason to the offset seen} in the posteriors of the {cosmological} parameters by {sampling instead} with priors centered around the {synthetic truth}. 
{We find that the resulting posteriors are close to previous results (grey vs. green in fig.~\ref{fig:synth})}, suggesting that the central value of the prior of the bias parameters does not play a substantial role. This means that even if the truth is the maximum likelihood point, the posteriors will not recover it.  

Having excluded that the bias in the posteriors on synthetic data is due to an error {in our pipeline} or due to our priors, we conclude that it must be due to phase-space projection effects. 
In fact, if the posteriors of the EFT parameters are effectively non-Gaussian ({\it i.e.} if the error bars are sufficiently large that the Taylor expansion {at second order} around the maximum of the posterior is not accurate enough to describe the actual posterior), then, upon marginalization, one can get projection effects on the remaining parameters, which, in this case, are the cosmological ones, even if the maximum likelihood point is the truth. Given the large number of EFT parameters, it is not so surprising that this might be the case. We call this effect `phase space effect,' but it is also known as `prior volume effect' or `projection effect.'

We decide to fix this issue with the following procedure. 
{As a measurement of the phase-space effect, for all analyses in this work, we take the shift in the 1D posteriors from the truth obtained fitting synthetic data with the same modeling and covariance. }
We add a prior of the following form to the $\log$-likelihood of $P_\ell + B_0 (+ B_2)$: 
\bea
&&\ln \calP_{\rm pr}^{\rm ph.\; sp.\; 1 sky}= -18 \left(\frac{b_1}{2}\right)+8 \left(\frac{\Omega_m}{0.31}\right)+14\left(\frac{h}{0.68}\right) \ ,\\ \nn
&&\ln \calP_{\rm pr}^{\rm ph.\; sp.\; 4 sky}= -48 \left(\frac{b_1}{2}\right)+32 \left(\frac{\Omega_m}{0.31}\right)+48\left(\frac{h}{0.68}\right)\ ,
\eea
respectively for 1 sky and 4 skies.\footnote{
When analyzing the power spectrum multipoles $P_\ell$ alone, we put the following prior instead: 
\bea
&&\ln \calP_{\rm pr}^{\rm ph.\; sp.\; 1 sky}= 2 \left(\frac{b_1}{2}\right)-2 \left(\frac{\Omega_m}{0.31}\right) \ ,\\ \nn
&&\ln \calP_{\rm pr}^{\rm ph.\; sp.\; 4 sky}= -4 \left(\frac{b_1}{2}\right)+10 \left(\frac{\Omega_m}{0.31}\right)+14\left(\frac{h}{0.68}\right)\ .
\eea }
As such, {upon marginalization, we recover unbiased 1D posteriors from the fit to the synthetic data} (see the red curve in fig.~\ref{fig:synth} and also the associated table).  
{More in detail, in tab.~\ref{table:synth}, we show the residual deviation from phase-space projection on the base cosmological parameters measured from synthetic data. 
We see that for all data volume (either the one of CMASS NGC or of all BOSS 4 skies) and cosmologies tested here (either the one of Nseries or the one of Patchy), we find that the residual deviation are negligibly small ($\lesssim 0.15$ or the error bars obtained with BOSS-volume covariance). }
Since the synthetic data are close to the Patchy ones (and so to the data), and since we expect the phase-space projection effects to be a slowly-varying function of the cosmological and EFT parameters, we add the same phase-space-correcting prior to the likelihood {of} the BOSS data.

\begin{table}[t]
  \centering
   \scriptsize
 \begin{tabular}{|l|c|c|c|c|c|c|} \hline
 $\Delta_{\rm sys}/\sigma_{\rm stat}$       & $\Omega_m$ & $h$  & $\sigma_8$ & $\omega_{cdm}$ & $\ln(10^{10}A_s)$ & $S_8$ \\ \hline
$P_\ell+B_0$: base - w/ NNLO & -0.03 & -0.09 & -0.03 &  -0.1 & 0.05 &  -0.04       \\ 
$P_\ell+B_0$: base - w/o $B_0$ window & 0.11 &  -0.05 &  0.01  & 0.05 & -0.01  &  0.05          \\ 
$P_\ell+B_0+B_2$: base - w/o $B_0, B_2$ window &  0.51 &  0.09 &  0.02 & 0.51 & -0.25  & 0.19         \\  \hline
\end{tabular}

  \caption{\footnotesize Relative shifts $\Delta_{\rm sys}/\sigma_{\rm stat}$ on base cosmological parameters measured from various modeling choices compared to our baseline: inclusion of the NNLO or removal of the window function in the bispectrum.  }
  \label{tab:sys}
  \end{table}

\subsection{Scale cut from NNLO}

A  simulation-independent way to evaluate the theoretical error as a function of $k_{\rm max}$ is to analyze the data by adding to the theory model a part of the next order terms: {for our one-loop model, this part consists in the next-to-next-to-leading-order} (NNLO) terms. Such a procedure was successfully applied to estimate the scale cut for the CF~\cite{Zhang:2021yna}. Here we use the same technique. We add the following two-loop counterterms to the EFTofLSS prediction at one-loop for the power spectrum:
\begin{equation}
  P_{\rm NNLO}(k, \mu) = \frac{1}{4}c_{r,4} b_1^2 \mu^4  \frac{k^4}{k_{\rm NL,R}^4} P_{11}(k) + \frac{1}{4}c_{r,6} b_1 \mu^6 \frac{k^4}{k_{\rm NL,R}^4} P_{11}(k)  \, ,
\end{equation}
and for the bispectrum: 
\begin{align}
&B_{\rm NNLO}(k_1, k_2, k_3, \mu, \phi) = 2 c_{\rm NNLO, 1}  K_2^{r,h} ( \kvec_1, \kvec_2 ;  \hat z ) K_1^{r,h} ( \kvec_2  ; \hat z )  f \mu_1^2 \frac{k_1^4}{k_{\rm NL, R}^4} P_{11}(k_1) P_{11}(k_2) \nonumber \\ 
&+ c_{\rm NNLO, 2} K_1^{r,h} ( \kvec_1  ; \hat z ) K_1^{r,h} ( \kvec_2  ; \hat z ) P_{11}(k_1) P_{11}(k_2) f  \mu_3 k_3 \frac{(k_1^2 + k_2^2)}{4 k_1^2 k_2^2 k_{\rm NL, R}^4}  \Big[- 2 \kvec_1 \cdot \kvec_2 (k_1^3 \mu_1 + k_2^3 \mu_2)  \nonumber \\
  &\quad + 2 f \mu_1 \mu_2 \mu_3 k_1 k_2 k_3   (k_1^2 + k_2^2)  \Big]  + \textrm{perm.} \, ,
\end{align}
where $k_{\rm NL, R} = k_{\rm NL}/\sqrt{8}$, as discussed in~sec.~\ref{sec:likelihood}. 
The prefactors {$c_{r,4}, c_{r,6}, c_{\rm NNLO, 1}$, and $c_{\rm NNLO, 2}$} are given a Gaussian prior centered on zero and {of} width 2. We then analyze the data as a function of $\kmax$, and determine the maximum wavenumber by taking the largest $\kmax$ where the shift {in all 1D posteriors of the cosmological parameters} with respect to the analysis without these terms is equal to $1/3\cdot\sigma$. This would mean that our results would have become sensitive to these terms that we do not fully compute, and so we need to analyze the data only up to this threshold. For simplicity, rather than determining the $\kmax$ in this way, we check the effect of these NNLO terms close to the $\kmax$ that we find in simulations, and check that the effect of the NNLO terms is indeed not too large. The results are presented in tab.~\ref{tab:sys}. We see that the effect is negligibly small, confirming what we find in simulations next, {\it i.e.} that our scale cut is appropriate.

\begin{figure}[t!]
  \centering
  \includegraphics[width=0.49\textwidth]{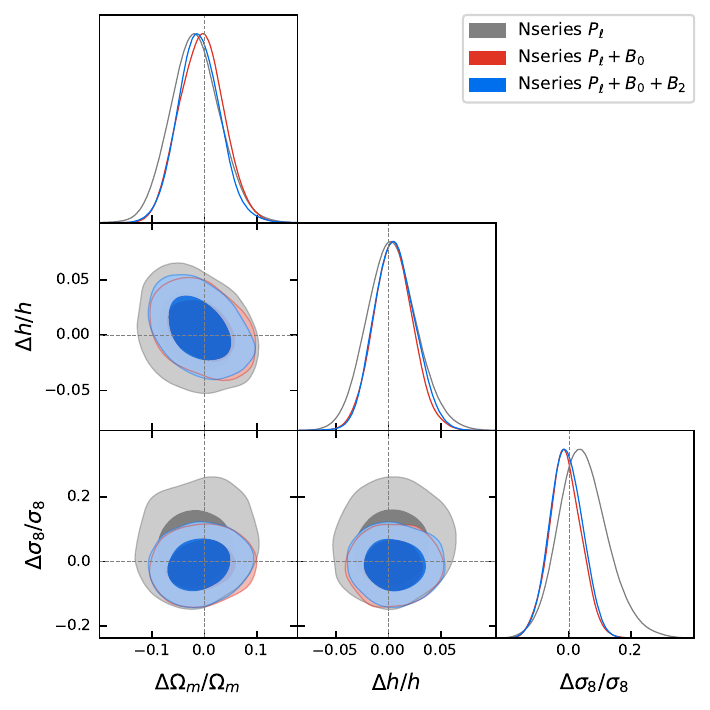}
  \includegraphics[width=0.49\textwidth]{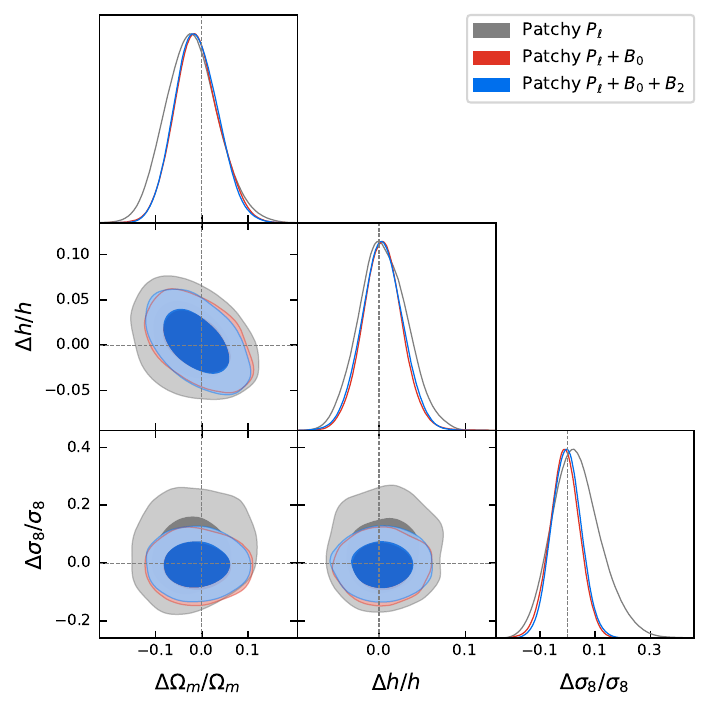}
    \vspace{0.5em}
    
\scriptsize
   \begin{tabular}{|p{0.7em}|p{5.5em}|p{6.5em}|p{5.5em}|p{6.5em}|p{6.5em}|p{6.5em}|p{6.5em}|}
     \hline 
 \multicolumn{2}{|c|}{$\Delta X / X$} 
 						& $\Omega_m$ 			&  $ h$					&  $\sigma_8 $ 					& $\omega_{cdm}$ 				& $\ln (10^{10}A_s)$ 		& $S_8$	\\ \hline 
 \multirow{3}{*}{\begin{turn}{90}Nseries\end{turn}}  
    & $P_\ell$ 				&  $-0.017\pm 0.048$		& $0.003^{+0.022}_{-0.024}$ 	& $0.047^{+0.070}_{-0.086}$ 		& $-0.013^{+0.063}_{-0.071}$ 		& $0.035\pm 0.055$ 			& $0.038^{+0.074}_{-0.092}$	\\
    & $P_\ell + B_0$ 			&  $-0.005\pm 0.042$		& $0.005\pm 0.019$ 			& $-0.012\pm 0.052$		 		& $0.004^{+0.052}_{-0.058}$ 		& $-0.010\pm 0.040$ 		& $-0.015\pm 0.058$	\\
    & $P_\ell + B_0 + B_2$ 	&  $-0.010\pm 0.041$		& $0.006^{+0.018}_{-0.021}$ 	& $-0.009\pm 0.053$ 			& $0.001\pm 0.053$ 				& $-0.007\pm 0.041$ 		& $-0.014\pm 0.059$	\\ \hline 
  \multirow{3}{*}{\begin{turn}{90}Patchy\end{turn} }  
    & $P_\ell$ 				&  $-0.021^{+0.052}_{-0.059}$	& $0.005\pm 0.028$ 			& $0.034^{+0.076}_{-0.098}$ 		& $-0.014\pm 0.078$ 			& $0.026^{+0.056}_{-0.066}$ 	& $0.022^{+0.083}_{-0.10}$		\\
    & $P_\ell + B_0$ 			&  $-0.011^{+0.044}_{-0.051}$	& $0.004\pm 0.023$ 			& $-0.011\pm 0.054$		 		& $-0.004^{+0.052}_{-0.062}$ 		& $-0.006\pm 0.044$ 		& $-0.017\pm 0.058$	\\
    & $P_\ell + B_0 + B_2$ 	&  $-0.012\pm 0.046$		& $0.004\pm 0.024$ 			& $-0.004\pm 0.053$ 			& $-0.006^{+0.052}_{-0.059}$ 		& $-0.001\pm 0.043$	 		& $-0.011\pm 0.058$	\\ 
    \hline 
    \end{tabular} \vspace{1em}

  \caption{\footnotesize 
   Triangle plots and relative $68\%$-credible intervals of base cosmological parameters measured from the Nseries and Patchy simulations analyzed using a covariance with CMASS NGC volume. 
  The grey lines in the triangle plots represent the simulation truth.  
  }
  \label{fig:simulations}
  \end{figure}

\subsection{Tests of additional modeling effects}

Our implementation of the IR-resummation and of the window function  is approximate, without a control parameter. We therefore check the accuracy of the two implementations in the following way.

For the window function, the correctness of our approximation has been checked in~\cite{Pardede:2022udo} for the monopole. 
In fact, as shown in the second line of tab.~\ref{tab:sys}, the difference between the bispectrum computed with our approximation, and the one where we apply no window is within $1/4$ of the error bars {obtained on all cosmological parameters from the fit to BOSS data}. For the quadrupole, the third line of tab.~\ref{tab:sys} shows that the difference with applying no window is about 0.5$\sigma$ on {the posterior of} $\Omega_m$ (while negligible for the other {cosmological parameters}). While this might seem too large an effect to tolerate, one should keep in mind the following. Roughly speaking, the correct window function should consist of applying $3/2$ factors of $W$ to the bispectrum ({\it i.e.} one for each field). Applying no window therefore is a radical negligence of all these factors, much worse than the approximation we do (which applies two factors of $W$). We therefore believe that a more reliable estimate of the error associated to our implementation of the window function for the quadrupole is obtained by dividing the effect in tab.~\ref{tab:sys} by a factor of 4. {Even if our estimate were to be wrong by a factor 2, } this {would} make the effect {safely} negligible.  {It would be interesting to compare our approach to an analysis using another estimator based on tri-polar spherical harmonics (described in \cite{Sugiyama:2018yzo} and tested on Patchy mocks in \cite{Sugiyama:2020uil}) for which the window functions can be estimated on an equal footing, making its application more straightforward. }

Let us now discuss the goodness of our approximate implementation of the IR-resummation of the bispectrum. It should be emphasized that the wiggle/no-wiggle procedure is affected by several uncontrolled approximations ({\it i.e.} not controlled by a small parameter, but numerically accidentally small)~\cite{Lewandowski:2018ywf}. On top of those, our formulas neglect the angle dependence of the IR-resummation, and, perhaps even more quantitatively importantly, do not damp the oscillations in the power spectra whose momenta are integrated in the loop integrals, as for example proposed in~\cite{Ivanov:2018gjr}. We checked that applying the damping for those power spectra leads to a negligible ($\lesssim 0.25$) change in the $\chi^2$ when keeping all the parameters of the model fixed.
{We therefore conclude that neglecting the IR-resummation on the `wiggly' parts from inside the loop integrals is accurate enough for BOSS data. 
We leave to future work more careful inspection of the remaining approximations in our IR-resummation scheme. }

\subsection{Tests against simulations}

\begin{table}[t]
\centering
\scriptsize
\begin{tabular}{|l|p{2em}|c|c|c|c|c|c|} \hline
 $\sigma_{\rm sys}^{\rm sim}/\sigma_{\rm stat}^{\rm data}$       & $\Omega_m$ & $h$  & $\sigma_8$ & $\omega_{cdm}$ & $\ln(10^{10}A_s)$ & $S_8$ \\ \hline
Nseries $P_\ell + B_0$ & 0.02 & 0.17 & 0.15 &-0.03  &0.17 &  0.17      \\ 
Nseries $P_\ell + B_0 + B_2$ & 0.16  & 0.25 & 0.08 & -0.09 & 0.08     & 0.16 \\ \hline
Patchy  $P_\ell + B_0$ & 0.27 & 0.21 & 0.23 & 0.05 & 0.14 & 0.33 \\
Patchy  $P_\ell + B_0 + B_2$ &  0.31 & 0.2 & 0.07 & 0.09 &  0   & 0.2 \\ \hline
\end{tabular} 
  \caption{\footnotesize 
   Report of systematic errors on base cosmological parameters measured from the Nseries and Patchy simulations. 
  The systematic error, reported relative to the BOSS error bars $\sigma_{\rm stat}^{\rm data}$, is defined as $\sigma_{\rm sys}^{\rm sim} \equiv \max(|\text{mean}-\textrm{truth}|- \sigma_{\rm stat}^{\rm sim} / \sqrt{N_{\rm sim}}, 0)$. 
 Here $\sigma_{\rm stat}^{\rm sim} / \sqrt{N_{\rm sim}}$ represents the uncertainty from the simulation cosmic variance, which corresponds to about $0.15$ or $0.03$ in $\sigma_{\rm stat}^{\rm data}$ for $N_{\rm sim} = 84$ Nseries or $N_{\rm sim} = 2048$ Patchy realizations, respectively. 
  }
  \label{tab:simulations2}
\end{table}

We {now} test the accuracy of the model by comparing against {$N$-body} simulations {described in sec.~\ref{sec:data}}. This does not only test  the effect of the theoretical error due to the next order terms {not included in our baseline model} or to the approximate IR-resummation, but also of the other observational effects that we model imperfectly, such as the window function. In fig.~\ref{fig:simulations}, we show the posterior{s from} the analysis of the average of 84 Nseries boxes, analyzed with the covariance of one box, {such that we also} account for the phase space effect. 
Since the actual cosmic variance associated to this average of 84 boxes is about 1/9 of the posterior{s} in fig.~\ref{fig:simulations}, we measure {for each cosmological parameter} the theoretical error as the distance of the mean of the posterior to the truth of the simulation minus 1/9 of the standard deviation (we take zero if this number is negative). 
This allows us to detect theoretical errors larger than 1/9 of a standard deviation of the posterior in fig.~\ref{fig:simulations}, which corresponds to about {0.15} of the {of the error bars obtained} on the BOSS data.
Our results show that the theoretical error that we can detect is safely below 1/3 of {the error bars obtained on BOSS}, as summarized in tab.~\ref{tab:simulations2}.

In {fig.~\ref{fig:simulations}}, we also present the analogous analysis on the average of 2048 Patchy mocks. 
In this case, the detectable theoretical error is {almost} unaffected by cosmic variance. Thus, assuming no systematic error in the Patchy simulations, the minimal detectable theoretical error is practically zero.
Also in this case, the theoretical error is safely below 1/3 {the error bars obtained on BOSS}, as summarized in tab.~\ref{tab:simulations2}. 

{We conclude that our analysis pipeline is free from significant systematics for BOSS volume at the scale cuts chosen at the beginning of this section, and we now move on to the analysis of the observational data.  }

%%%%%%%%%%%%%%%%%%%%%%%%%%%%%%%%%%%%%%%%%%%%%%%%%%%%%%%%%%%%%
%%%%%%%%%%%%%%%%%%%%%%%%%%%%%%%%%%%%%%%%%%%%%%%%%%%%%%%%%%%%%

\section{Results} \label{sec:results}

\begin{table}[t!]
\centering
\scriptsize
\begin{tabular}{|l|c|c|c|c|} \hline
 & $N_{\rm bin}$ // dof & $\min \chi^2$ & $\min \chi^2/$dof & $p$-value \\ \hline
CMASS NGC  & $42+150+9=201$ & 159.5 & 0.79 & 0.99\\ 
CMASS SGC  & $42+150+9=201$ & 188.7 & 0.94 & 0.72\\ 
LOWZ NGC  & $36+62+9=107$ & 98.3 & 0.92 & 0.71\\ 
LOWZ SGC  & $36+62+9=107$ & 106.4 & 0.99 & 0.50\\ 
Parameter Prior & {$3+41 (1+0.1 + 0.2 + 0.1 \cdot 0.2) \simeq 57$} & 8.9 & - & - \\
Total & {$616-57=559$ }&  $561.9$ &  {1.01 } & {0.46} \\ \hline
\end{tabular} 
  \caption{\footnotesize 
Goodness of fit given by the maximal $\log$-likelihood value $\log \mathcal{L} \equiv - \min \chi^2 / 2$ obtained fitting BOSS 4 skies $P_\ell + B_0 + B_2$, and associated $p$-value. 
For each skycut, we detail the number of bins $N_{\rm bin} = N_{\rm bin}^{P_\ell} + N_{\rm bin}^{B_0} + N_{\rm bin}^{B_2}$, while in `Parameter Prior' we give instead the degrees of freedom (dof). 
The dof are {taken as} the sum of 3 varied cosmological parameters (that are not prior dominated) plus an effective number of correlated EFT parameters. 
The $p$-value are calculated assuming there is no correlation within the data. 
  }
  \label{tab:bestfit}
\end{table}

\begin{figure}[t!]
  \centering
  \includegraphics[width=0.67\textwidth]{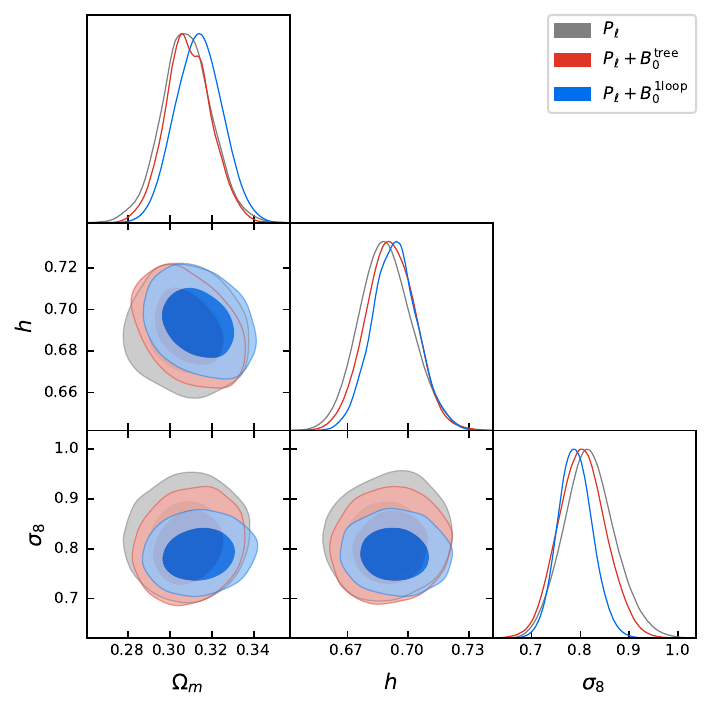}
      \caption{\footnotesize
{Triangle plots of base cosmological parameters measured from the analysis of BOSS power spectrum multipoles $P_\ell$, $\ell=0,2$, at one-loop, and bispectrum monopole $B_0$ at tree or one-loop level. } }
  \label{fig:boss_b0}
  \end{figure}

When analyzing the BOSS data, we find that there is no additional gain by adding all the three independent quadrupoles after one has been included. We therefore present results including only {$B_{(2,3)}^{r,h}$}.

In fig.~\ref{fig:bestfit}, we show the best fit residuals and in tab.~\ref{tab:bestfit} the best-fit $\chi^2$ and associated $p$-value. The $p$-value is very good and we do not find any concerning systematic behavior in the residuals. In fig.~\ref{fig:boss}, we provide the best-fit parameters, which safely lie within our $68\%$-credible intervals. 

The posteriors associated to the analysis of the BOSS data are presented in fig.~\ref{fig:boss} {and fig.~\ref{fig:boss_b0}}. 
They are discussed in the Introduction. 
In App.~\ref{app:constraints} we provide the posteriors for the other non-marginalized parameters as well their confidence interval. One can see that the bispectrum improves their measurement by order 100\%.

%%%%%%%%%%%%%%%%%%%%%%%%%%%%%%%%%%%%%%%%%%%%%%%%%%%%%%%%%%%%%

%%%%%%%%%%%%%%%%%%%%%%%%%%%%%%%%%%%%%%%%%%%%%%%%%%%%%%%%%%%%%

\section*{Acknowledgements}

\noindent  We thank {Babis Anastasiou, Diogo Bracanca and Henry Zheng for letting us use the code to evaluate the loop integrals in~\cite{Anastasiou:2022udy}.}
We thank Matteo Biagetti, Hector Gil-Mar\'in, Emiliano Sefusatti, and Cheng Zhao for useful discussions. 
M.L. acknowledges the Northwestern University Amplitudes and Insight group, Department of Physics and Astronomy, and Weinberg College, and is also supported by the DOE under contract DE-SC0021485. 
P.Z. is grateful to Yi-Fu Cai for support. 
Y.D. acknowledges support from the STFC.
Part of the analysis was performed on the HPC (High Performance Computing) facility of the University of Parma, whose support team we thank, part on the HPC environment in CINECA thanks to the InDark project, and part on the computer clusters LINDA $\&$ JUDY in the particle cosmology group at USTC.

%%%%%%%%%%%%%%%%%%
%

\appendix
\section{EFTofLSS details}\label{theoryapp}

%%%%%%%%%%%%
%
%
\subsection{General expressions}

Here we give the details necessary to compute the one-loop power spectrum and one-loop bispectrum of biased tracers in redshift space in the EFTofLSS.  The bias expansion for the halo overdensity is given by 
		\begin{align}
		\begin{split} \label{finalbias}
		\delta_{h}(\vx, t ) = & b_1 \left(\mathbb{C}_{\delta,1}^{(1)}(\vx , t )+\mathbb{C}_{\delta,1}^{(2)}(\vx , t )+\mathbb{C}_{\delta,1}^{(3)}(\vx , t )+\mathbb{C}_{\delta,1}^{(4)}(\vx , t )\right)  \\
		& +  b_2 \left(\mathbb{C}_{\delta,2}^{(2)}(\vx , t )+\mathbb{C}_{\delta,2}^{(3)}(\vx , t )+\mathbb{C}_{\delta,2}^{(4)}(\vx , t )\right)  +  b_3  \left(\mathbb{C}_{\delta,3}^{(3)}(\vx , t )+\mathbb{C}_{\delta,3}^{(4)}(\vx , t )\right) \\
		& + b_4 \;\mathbb{C}_{\delta,4}^{(4)}(\vx , t )  +  b_5  \left(\mathbb{C}_{\delta^2,1}^{(2)}(\vx , t )+\mathbb{C}_{\delta^2,1}^{(3)}(\vx , t )+\mathbb{C}_{\delta^2,1}^{(4)}(\vx , t )\right)\\ 
		& + b_6 \left(\mathbb{C}_{\delta^2,2}^{(3)}(\vx , t )+\mathbb{C}_{\delta^2,2}^{(4)}(\vx , t )\right) +  b_7   \,  \mathbb{C}_{\delta^2,3}^{(4)}(\vx , t ) + b_8 \left(\mathbb{C}_{r^2,2}^{(3)}(\vx , t )+\mathbb{C}_{r^2,2}^{(4)}(\vx , t )\right)\\
		& + b_9 \, \mathbb{C}_{r^2,3}^{(4)}(\vx , t ) + b_{10}  \left(\mathbb{C}_{\delta^3,1}^{(3)}(\vx , t )+\mathbb{C}_{\delta^3,1}^{(4)}(\vx , t )\right)+ b_{11} \, \mathbb{C}_{r^3,2}^{(4)}(\vx , t ) \\ 
		&+ b_{12} \,  \mathbb{C}_{\delta^3,2}^{(4)}(\vx , t )+ b_{13} \,  \mathbb{C}_{r^2\delta,2}^{(4)}(\vx , t )+ b_{14} \, \mathbb{C}_{\delta^4,1}^{(4)}(\vx , t )+ b_{15}\,  \mathbb{C}_{\delta r^3,1}^{(4)}(\vx , t )  \ .  
		\end{split}
		\end{align}
In the above, the $\mathbb{C}^{(n)}_{\mathcal{O},\alpha}$ functions are defined by Taylor expanding the operator $\mathcal{O}$ in the fluid line element $\xvec_{\rm fl}$ {\cite{Senatore:2014eva}}, which is given recursively by
\be
\xvec_{\rm fl} ( \xvec , t , t' ) = \xvec - \int_{t'}^t \frac{dt''}{a(t'')} \vec{v} ( \xvec_{\rm fl}( \xvec , t , t'') , t'' )  \ .
\ee
Explicitly, writing $\mathcal{O}_{m}$ to represent an operator that is the product of $m$ powers of fluctuations {(i.e. $m = 3$ for $\delta^3$,  $\partial^2 \phi \partial_i \partial_j \phi \partial^i \partial^j \phi$, etc.)}, we have
\be \label{biasfnbasis}
[ \mathcal{O}_{m} ( \xvec_{\rm fl} ( t , t') , t') ]^{(n)} =   \sum_{\alpha = 1}^{n-(m-1)}  \left( \frac{D(t' )}{D(t)} \right)^{\alpha+m-1}  \mathbb{C}^{(n)}_{\mathcal{O}_m , \alpha}  ( \xvec , t )  \ , 
\ee
where the notation $[\dots]^{(n)}$ means to take the $n$-th order term in the perturbative expansion.   Explicit expressions for the operators up to third order can be found in \cite{Fujita:2016dne}, with the identification $\mathbb{C}^{(3)}_{r^2,2} = \mathbb{C}^{(3)}_{s^2 , 2} + \mathbb{C}^{(3)}_{\delta^2 , 2} / 3 $, and fourth order operators can be found with \eqn{biasfnbasis}.  Operators at fourth order that are not related to previously used cubic terms are
\begin{align}
\begin{split}
&  \mathbb{C}^{(4)}_{r^3,2} ( \xvec , t )  = [ r^3( \xvec , t) ]^{(4)} -  \partial_i r^3( \xvec , t) \frac{\partial_i \theta ( \xvec , t) }{\partial^2 }  \ ,  \\ 
& \mathbb{C}^{(4)}_{r^2 \delta , 2 } ( \xvec , t ) = [ r^2 ( \xvec , t) \delta ( \xvec , t) ]^{(4)}  - \partial_i ( r^2 ( \xvec , t ) \delta( \xvec, t ) ) \frac{\partial_i \theta( \xvec , t) }{\partial^2} \ ,  \\
& \mathbb{C}^{(4)}_{\delta^4,1} ( \xvec , t  ) = \delta(\xvec , t )^4 \ , \quad \mathbb{C}^{(4)}_{\delta r^3,1} ( \xvec , t ) = \delta( \xvec , t) r^3 ( \xvec , t) \ ,
\end{split}
\end{align}
where 
 \be
r_{ij} \equiv  \partial_i \partial_j \phi  \ , \quad r^2 \equiv r_{ij} r_{ji} \ , \quad r^3 \equiv r_{ij}r_{jk}r_{ki} \ , \quad \partial^2 \phi \equiv \delta \andd \theta \equiv - \frac{\partial_i v^i}{f a H}    \ . 
\ee

  Given the position space halo overdensity $\delta_h $, the transformation to redshift space, $\delta_{r,h}$, is accomplished by
\begin{align}
\begin{split} \label{rsssbias}
\delta_{r,h}  &  = \delta_h  - \frac{\hat{z}^i \hat{z}^j}{aH } \partial_i \left( ( 1 + \delta_h ) v^j \right) + \frac{\hat{z}^i \hat{z}^j \hat{z}^k \hat{z}^l}{2 (a H)^2  } \partial_i \partial_j ( ( 1 + \delta_h ) v^k  v^l ) \\
& - \frac{\prod_{a=1}^6 \hat{z}^{i_a} }{3! (a H)^3 } \partial_{i_1} \partial_{i_2} \partial_{i_3} (( 1 + \delta_h ) v^{i_4} v^{i_5} v^{i_6} ) + \frac{\prod_{a=1}^8 \hat{z}^{i_a} }{4! (a H)^4} \partial_{i_1} \partial_{i_2} \partial_{i_3}\partial_{i_4}  (  v^{i_5 }  v^{i_6} v^{i_7} v^{i_8} )  + \dots \ ,
\end{split}
\end{align}
where $\hat z $ is the line-of-sight direction.  We then expand the overdensity perturbatively as
\be
\delta_{r,h} ( \kvec ; \hat z ) = \sum_n  \delta^{(n)}_{r,h} ( \kvec ; \hat z) \ .
\ee
For the one-loop power spectrum and the {one-loop} bispectrum, we need the overdensity to {fourth} order, $n=4$.  The solutions can be written as an expansion in powers of the linear dark-matter overdensity $\delta^{(1)}$ in terms of the symmetric $n$-th order halo kernels $K_{n}^{r,h}$ defined as in\footnote{We have introduced the following notation
\be
\int_{\kvec_1 , \dots , \kvec_n} \equiv \int \frac{\rmd^3 k_1 }{(2 \pi)^3} \cdots \frac{
\rmd^3 k_n}{(2 \pi)^3 } \ ,  \quad \int_{\kvec_1 , \dots , \kvec_n}^{\kvec} \equiv \int_{\kvec_1 , \dots , \kvec_n} ( 2 \pi)^3 \delta_D ( \kvec - \sum_{i = 1}^n \kvec_i )  \ .
\ee
{Additionally, $P_{11}$ is the dark-matter linear power spectrum.}}
\begin{align}
\begin{split} \label{halorsskernels}
\delta_{r,h}^{(1)} ( \kvec ; \hat z ) & = K_1^{r,h} ( \kvec ; \hat z ) \delta^{(1)} ( \kvec ) \ ,  \\
\delta_{r,h}^{(n)} ( \kvec  ; \hat z ) & =  \int_{\kvec_1 , \dots , \kvec_n}^{\kvec } K_n^{r,h} ( \kvec_1 , \dots , \kvec_n  ; \hat z ) \delta^{(1)} ( \kvec_1 ) \cdots \delta^{(1)}(\kvec_n) \ , \quad \text{for } n \geq 2 \ . 
\end{split}
\end{align}
For example,
\be
 K_1^{r,h} ( \kvec ; \hat z )  =  b_1 + f (\hat k \cdot \hat z )^2    \ , 
\ee
is the famous Kaiser result for linear theory.  The explicit expressions for $K_2^{r,h} $ and $K_3^{r,h} $ are given in \cite{Perko:2016puo}, while the expression for $K_4^{r,h}$ is available in \cite{DAmico:2022ukl} {and can be straightforwardly derived from the above expressions}.  We provide the dependence of the kernels $K_{1,\dots,4}^{r,h}$  on the bias parameters $\{ b_i \}$ for $i = 1, \dots, 15$ here for convenience 
\begin{align}
\begin{split} \label{kernelbiasdep}
& K_{1}^{r,h} [ b_1 ] \ , \quad  K_{2}^{r,h} [ b_1 , b_2 , b_5 ] \ , \quad  K_{3}^{r,h} [ b_1 , b_2 , b_3, b_5 , b_6, b_8, b_{10}  ]  \andd  K_4^{r,h} [ b_1, \dots , b_{15}]  \  . 
\end{split}
\end{align}

Given the perturbative expansion above, we can write the observables of interest (in our case the one-loop power spectrum and the one-loop bispectrum) in terms of the kernels $K^{r,h}_n$.  The relevant quantities for the power spectrum that enter \eqn{psexpansion} are the tree-level power spectrum
\be 
P_{11}^{r,h} ( k , \hat k \cdot \hat z) = (b_1 + f (\hat k \cdot \hat z )^2  )^2  P_{11} ( k )  \ , 
\ee
and the one-loop contributions
\begin{align}
\begin{split} \label{loopexpressionsrssbias}
&P_{22}^{r,h} ( k, \hat k \cdot \hat z  )  = 2 \int_{\qvec} K_2^{r,h} ( \qvec , \kvec - \qvec ; \hat z )^2 P_{11} ( q ) P_{11}( | \kvec - \qvec|) \ , \\
& P_{13}^{r,h}  (k, \hat k \cdot \hat z)  = 6 P_{11} ( k )K_1^{r,h} ( \kvec ; \hat z )  \int_{\qvec} K_3^{r,h} ( \qvec , - \qvec , \kvec ;  \hat z ) P_{11}(q) \ .
\end{split}
\end{align}
For the bispectrum, the quantities that enter \eqn{bispexpansion} are the tree-level bispectrum
\be
B_{211}^{r,h} = 2  K_1^{r,h} ( \kvec_1 ; \hat z )  K_1^{r,h} ( \kvec_2  ; \hat z )  K_2^{r,h} ( -\kvec_1, -\kvec_2 ;  \hat z ) P_{11} ( k_1 ) P_{11} ( k_2 ) + \text{ 2 perms.}  \ ,
\ee
and the one-loop contributions
\begin{align}
 \label{bispexpressionsrssbias}
& B^{r,h}_{222}  = 8 \int_{\qvec} P_{11}(q) P_{11}(| \kvec_2 - \qvec| ) P_{11} ( |\kvec_1 + \qvec|) \nonumber \\
& \hspace{1.5in} \times K_2^{r,h} ( - \qvec , \kvec_1 + \qvec ; \hat z) K^{r,h}_2 ( \kvec_1 + \qvec , \kvec_2 - \qvec ; \hat z ) K^{r,h}_2 ( \kvec_2 - \qvec , \qvec ;  \hat z) \ , \nonumber   \\
& B_{321}^{r,h,(I)}  = 6 P_{11}(k_1) K_1^{r,h} ( \kvec_1 ;  \hat z)  \int_{\qvec} P_{11}(q) P_{11}(| \kvec_2 - \qvec|)      \\
& \hspace{1.5in} \times K_3^{r,h} ( - \qvec , - \kvec_2 + \qvec , - \kvec_1 ; \hat z ) K^{r,h}_2 ( \qvec , \kvec_2 - \qvec  ;  \hat z )  + \text{ 5 perms.}\ ,   \nonumber   \\
& B_{321}^{r,h,(II)}  = 6 P_{11}(k_1 ) P_{11}(k_2) K_1^{r,h} ( \kvec_1  ; \hat z) K_2^{r,h} ( \kvec_1 , \kvec_2 ;  \hat z ) \int_{\qvec} P_{11}(q) K_3^{r,h} ( \kvec_2 , \qvec , - \qvec ; \hat z ) + \text{ 5 perms.} \ ,  \nonumber \\ 
& B_{411}^{r,h}  = 12 P_{11} ( k_1 ) P_{11}(k_2) K_1^{r,h} ( \kvec_1 ; \hat z) K_1^{r,h} ( \kvec_2 ; \hat z)  \int_{\qvec} P_{11}(q) K_4^{r,h} ( \qvec , - \qvec , - \kvec_1 , - \kvec_2 ; \hat z ) + \text{ 2 perms.}  \ . \nonumber
\end{align}
Note that on the left-hand sides we have dropped the argument $( k_1 , k_2 , k_3 , \hat k_1 \cdot \hat z , \hat k_2 \cdot \hat z ) $ on the bispectrum expressions to remove clutter.

Next we move on to the EFT counterterm contributions where, for renormalization up to the one-loop bispectrum, we need the EFT counterterms to second order in fields.  
Additionally, since we work with biased tracers, we can introduce the counterterms directly into \eqn{rsssbias}, i.e. directly at the level of biased tracers in redshift space, as in \cite{Perko:2016puo,Lewandowski:2015ziq} for example.  
We divide the counterterm contributions into two sources, response terms which are proportional to powers of the linear field $\delta^{(1)}$, and stochastic terms which contain randomly fluctuating fields, typically denoted with an `$\epsilon$.'  We can write the response terms as 
\begin{align}
\begin{split} \label{biasctexp}
& \delta^{(1)}_{r,h,ct}(\kvec ; \hat z  ) =   K_1^{r,h,ct} ( \kvec ;\hat z) \deltaone ( \kvec )  \ ,  \\ 
& \delta^{(2)}_{r,h,ct} ( \kvec ; \hat z  ) =  \int_{\qvec_1 , \qvec_2}^{\kvec} K_2^{r,h,ct} ( \qvec_1 , \qvec_2 ; \hat z  ) \deltaone ( \qvec_1 ) \deltaone ( \qvec_2 ) \  ,
\end{split}
\end{align}
where
\be
K_1^{r,h,ct} ( \kvec ; \hat z)  = \frac{k^2}{\knl^2} \left(  - c_{h,1} + f (\hat k \cdot \hat z)^2 c_{\pi,1} - \half f^2 (\hat k \cdot \hat z)^4 c_{\pi v , 1} - \half f^2 (\hat k \cdot \hat z)^2 c_{\pi v,3} \right) \ , 
\ee
and 
\begin{align}
\begin{split} \label{k2rhct}
K_2^{r,h,ct} ( \kvec_1 , \kvec_2 ; \hat z ) = \sum_{i=1}^{14} c_i   \, e^{K_2}_i ( \kvec_1 , \kvec_2 ; \hat z)  \ ,
\end{split}
\end{align}
with
\be
c_i = \{  c_{h,1} , c_{h,2} ,c_{h,3} ,c_{h,4} ,c_{h,5} , c_{\pi,1} , c_{\pi,5} , c_{\pi v ,1} , c_{\pi v ,2} , c_{\pi v ,3} , c_{\pi v ,4} , c_{\pi v ,5} , c_{\pi v ,6} , c_{\pi v ,7} \} \ ,
\ee
and the $e^{K_2}_i $ functions are {given below} in \secref{functionssec}.  Similarly, we denote the first order stochastic term as $ \delta_1^{r,h,\epsilon} ( \kvec ; \hat z )$, and the second order ones as
\begin{align}
\begin{split}
& \delta^{(2)}_{r,h,\epsilon} ( \kvec ; \hat z  ) =  \int_{\qvec_1 , \qvec_2}^{\kvec} \delta_2^{r,h,\epsilon} ( \qvec_1 , \qvec_2; \hat z) \delta^{(1)} ( \qvec_2 ) \ .
\end{split}
\end{align}
Here, $\delta_1^{r,h,\epsilon} ( \kvec ; \hat z )$ and $\delta_2^{r,h,\epsilon} ( \qvec_1 , \qvec_2; \hat z)$ contain all allowed contractions of tensor stochastic fields $\epsilon^{ij\dots}$, which we assume to be Poisson distributed and do not correlate with the matter field $\delta^{(1)}$.  Contractions of the stochastic fields are defined as in \cite{Lewandowski:2015ziq}.  

In terms of the kernels above, the response counterterms are
\begin{align}
\begin{split}
&P_{13}^{r,h,ct} ( k , \hat k \cdot \hat z ) = 2 K_1^{r,h}( \kvec ; \hat z ) K_1^{r,h,ct} ( - \kvec ; \hat z ) P_{11} ( k ) \  , \\
& B_{321}^{r,h,(II),ct}  = 2 P_{11} ( k_1 ) P_{11} ( k_2 ) K_{1}^{r,h,ct} ( \kvec_1 ;  \hat z ) K_1^{r,h} ( \kvec_2 ; \hat z ) K^{r,h}_2 ( - \kvec_1 , - \kvec_2 ; \hat z )  + \text{ 5 perms.} \ , \\
& B_{411}^{r,h,ct}  = 2 P_{11} ( k_1 ) P_{11} ( k_2 )K_1^{r,h}(\kvec_1 ; \hat z ) K_1^{r,h} (\kvec_2 ; \hat z )   K_{2}^{r,h,ct} ( - \kvec_1 , - \kvec_2 ; \hat z  )  + \text{ 2 perms.} \ .
\end{split}
\end{align}
The two stochastic contributions that involve only $\delta_1^{r,h,\epsilon}$ and do not contain any long-wavelength fields are
\be
P_{22}^{r,h,\epsilon}  = \frac{1}{\bar n} \left( c_1^{\rm St}  +c^{\rm St}_2  \frac{k^2}{\knl^2}  + c^{\rm St}_3 \frac{k^2}{\knl^2} f ( \hat k \cdot \hat z)^2    \right)  \ , 
\ee
and
\be
B_{222}^{r,h,\epsilon}  =\frac{1}{\bar n^2}  \left( c_1^{(222)} + \frac{1}{\knl^2} \left( c^{(222)}_2 (k_1^2 + k_2^2 + k_3^2 )   + c_5^{(222)} \hat z^i \hat z^j \left(k_1^i k_2^j + k_1^i k_3^j + k_2^i k_3^j  \right) \right) \right) \ .
\ee 
The final, mixed response-stochastic, contribution is slightly more complicated, so we first define $ \tilde B_{321}^{r,h, (I), \epsilon} $ from 
\begin{align}
\begin{split} \label{btilde321}
   & ( 2 \pi)^3 \delta_D ( \kvec_1 + \kvec_2 + \kvec_3 ) \tilde B_{321}^{r,h, (I), \epsilon} (\vec  k_1 ,\vec k_2 ,  \vec k_3 ; \hat z) = \\
 &  \hspace{1in} \langle \delta_{r,h}^{(1)} ( \kvec_1 ; \hat z  ) \delta_1^{r,h,\epsilon} ( \kvec_2 ; \hat z ) \delta^{(2)}_{r,h,\epsilon} ( \kvec_3 ; \hat z ) \rangle + \langle \delta_{r,h}^{(1)} ( \kvec_1 ; \hat z  ) \delta_1^{r,h,\epsilon}  ( \kvec_3 ; \hat z ) \delta^{(2)}_{r,h,\epsilon} ( \kvec_2 ; \hat z ) \rangle   \ ,
\end{split}
\end{align}
so that 
\be
B_{321}^{r,h, (I), \epsilon}  = \tilde B_{321}^{r,h, (I), \epsilon} ( \vec k_1 , \vec k_2 , \vec k_3 ;  \hat z) +\tilde B_{321}^{r,h, (I), \epsilon} ( \vec k_3 , \vec k_1 , \vec k_2 ; \hat z )  + \tilde B_{321}^{r,h, (I), \epsilon} ( \vec k_2 , \vec k_3 , \vec k_1  ;  \hat z)   \ .
\ee
Then we have
\be \label{b321ctexp}
\tilde B_{321}^{r,h, (I), \epsilon} ( \vec k_1 ,  \vec k_2 ,  \vec k_3 ; \hat z ) = \frac{( b_1 + f ( \hat k_1 \cdot \hat z)^2)}{\bar n} P_{11}(k_1) \sum_{i=1}^{13}   c^{\rm St}_i  e_i^{\rm St} (\vec k_1 ,   \vec k_2 , \vec  k_3 ; \hat z)  \  .
\ee
The $ e_i^{\rm St} $ functions are {given below} in \secref{functionssec}. 
Notice that $e_3^{\rm St} = 0$, so that there is no $c_3^{\rm St}$ parameter.  

{A full description of the fourth order bias expansion and renormalization in redshift space is given in \cite{DAmico:2022ukl}.  See \cite{Eggemeier:2021cam} for the bias expansion for the real space one-loop bispectrum (and measurement of the scalar amplitude $A_s$ from simulations).}

%%%%%%%%%%%%
%
%
\subsection{Explicit functions} \label{functionssec}
The functions that enter $K_2^{r,h,ct}$ in \eqn{k2rhct} are given by  
\begin{align}
\begin{split}
e_1^{K_2} & = \frac{-2 f k_2 k_1^3 \mu _1 \mu _2-2 f k_2^4 \mu _1^2+k_1^4+\left(k_2^2-k_3^2\right) k_1^2}{4 k_2^2
   k_{\text{NL}}^2} + (1 \leftrightarrow 2)  \ ,  \\
e_2^{K_2} & = - \frac{1}{28 k_1^2 k_2^2 k_{\text{NL}}^2}  \left( k_3^6+5 k_1^2 k_2^2 k_3^2  +  7 k_1^6+\left(7 k_2^2-12 k_3^2\right) k_1^4+3 k_2^2 k_3^4 + ( 1 \leftrightarrow 2) \right)   \ , \\
   e_3^{K_2} & =  - \frac{k_3^2}{ \knl^2 } \ , \quad    e_4^{K_2}  =     -\frac{k_3^2 \left(k_1^2+k_2^2-k_3^2\right){}^2}{4 k_1^2 k_2^2 k_{\text{NL}}^2}   \ , \quad e_5^{K_2}  =     \frac{k_1^2+k_2^2-k_3^2}{2 k_{\text{NL}}^2}   \ , \\ 
   \end{split}
\end{align}

\begin{align}
e_6^{K_2} & = \frac{f (k_1 \mu_1 + k_2 \mu_2)}{4  k_1^2 k_2^2 k_{\text{NL}}^2} \left(   k_1^2 \mu _1 \left(2 f k_2 k_1^2 \mu _1 \mu _2+2 f k_2^3 \mu _1 \mu _2-k_1^3+\left(k_3^2-k_2^2\right) 
   k_1\right)  + ( 1 \leftrightarrow 2)  \right) \nonumber  \ ,  \\
   e_7^{K_2} & =   \frac{f \left(k_1^2-k_2^2-k_3^2\right) \left(k_1^2+k_2^2-k_3^2\right) \left(k_1^2-k_2^2+k_3^2\right) \left(k_1 \mu _1+k_2 \mu _2\right){}^2}{8 k_1^2
   k_2^2 k_3^2 k_{\text{NL}}^2}  \ ,      \\
   e_8^{K_2} & =   \frac{f^2 (k_1 \mu_1 + k_2 \mu_2)^2}{8  k_1^2 k_2^2  k_{\text{NL}}^2 } \left( - 2 f k_1^2 k_2^2 \mu _1^2 \mu _2^2  +   k_1^2 \mu _1^2 \left(-2 f k_2 k_1 \mu _1 \mu _2+k_1^2+k_2^2-k_3^2\right) + ( 1 \leftrightarrow 2)      \right) \ ,   \nonumber
\end{align}

\begin{align}
\begin{split}
e_9^{K_2} & = \frac{f^2 (k_1 \mu_1 + k_2 \mu_2)^2}{56  k_1^2 k_2^2 k_3^2 k_{\text{NL}}^2 } \Bigg( -2 k_1 k_2 \left(k_3^4+5 k_1^2 k_2^2\right) \mu _1 \mu _2 +    k_1^2 \mu _1 \Big( 5 k_1^4 \mu _1+10 k_2 k_1^3 \mu _2 \\
&  \hspace{1in}  -10 \left(k_2^2+k_3^2\right) k_1^2 \mu _1-6
   k_2 k_3^2 k_1 \mu _2+5 \left(k_2^2-k_3^2\right){}^2 \mu _1 \Big) + ( 1 \leftrightarrow 2)  \Bigg)  \ , \\
e_{10}^{K_2} & =  \frac{f^2 (k_1 \mu_1 + k_2 \mu_2 )^2}{8  k_1^2 k_2^2 \knl^2 }  \left( k_1^2 k_2^2 + k_1^2 \left(-2 f k_2 k_1 \mu _1 \mu _2-2 f k_2^2 \mu _1^2+k_1^2-k_3^2\right) + ( 1 \leftrightarrow 2)       \right) \ ,  \\
   e_{11}^{K_2} & =  \frac{f^2 (k_1 \mu_1 + k_2 \mu_2 )^2}{28  k_1^2 k_2^2 \knl^2 }   \left( -k_3^4-12 k_1^2 k_2^2 + \left(2 k_3^2-k_1^2   \right)k_1^2 + \left(2 k_3^2-k_2^2   \right)k_2^2  \right)  \ ,   \\
   e_{12}^{K_2} & =    -\frac{f^2 \left(\mu _1^2+\mu _2^2\right) \left(k_1 \mu _1+k_2 \mu _2\right){}^2}{4 k_{\text{NL}}^2}   \ , \quad  e_{13}^{K_2}   =    \frac{f^2 \left(k_1^2+k_2^2-k_3^2\right) \mu _1 \mu _2 \left(k_1 \mu _1+k_2 \mu _2\right){}^2}{4 k_1 k_2 k_{\text{NL}}^2}  \ ,    \\
     e_{14}^{K_2} & =     -\frac{f^2 \left(k_1 \mu _1+k_2 \mu _2\right){}^2}{2 k_{\text{NL}}^2}  \ . 
   \end{split}
\end{align}
In the above, we have used the notation $\mu_i \equiv \hat k_i \cdot \hat z$.  

The functions that enter the stochastic counterterm $B_{321}^{r,h,(I),\epsilon}$ in \eqn{b321ctexp} are:
\begin{align}
& e_1^{\rm St} =   f \mu _1^2-1 \ ,   \nonumber  \\
& e_2^{\rm St} = -\frac{k_1^2 \left(k_2^2 \left(1-2 f \mu _1^2\right)+k_3^2\right)+2 f k_2 \left(k_3^2-k_2^2\right) k_1 \mu _1
   \mu _2+\left(k_2^2-k_3^2\right){}^2}{2 k_1^2 k_{\text{NL}}^2}  \ ,  \\
  & e_3^{\rm St} = 0 \ , \nonumber \\
& e_4^{\rm St} = -\frac{f^2 \mu _1 \left(k_1^3 \mu _1 \left(2 f \mu _1^2-1\right)+4 f k_2 k_1^2 \mu _1^2 \mu _2+k_1 \mu _1
   \left(k_2^2 \left(4 f \mu _2^2-1\right)+k_3^2\right)+2 k_2 \left(k_3^2-k_2^2\right) \mu _2\right)}{4 k_1
   k_{\text{NL}}^2}   \ ,   \nonumber  \\
& e_5^{\rm St} =  \frac{f^2 \mu _1 \left(4 f k_2 k_1^2 \mu _1^2 \mu _2+k_1 \mu _1 \left(k_2^2 \left(4 f \mu
   _2^2-1\right)+k_3^2\right)+k_1^3 \mu _1+2 k_2 \left(k_3^2-k_2^2\right) \mu _2\right)}{4 k_1
   k_{\text{NL}}^2}   \ , \nonumber  
   \end{align}
   \begin{align}
& e_6^{\rm St} =    2   \ , \quad  e_7^{\rm St} =    -\frac{k_2^2+k_3^2}{k_{\text{NL}}^2}     \ , \quad e_8^{\rm St} =    -\frac{k_1^4+\left(k_2^2-k_3^2\right){}^2}{2 k_1^2 k_{\text{NL}}^2}   \ , \nonumber   \\
& e_9^{\rm St} =    -\frac{k_1^2}{k_{\text{NL}}^2}   \ , \quad e_{10}^{\rm St} =    -\frac{f \left(k_1 \mu _1+2 k_2 \mu _2\right) \left(\left(k_1^2-k_2^2+k_3^2\right) \mu _1+2 k_1 k_2 \mu _2\right)}{4 k_1 k_{\text{NL}}^2}    \ ,  \\
& e_{11}^{\rm St} =     \frac{f \mu _1 \left(k_1 \left(k_1^2+k_2^2-k_3^2\right) \mu _1+2 k_2 \left(k_2^2-k_3^2\right) \mu _2\right)}{2 k_1 k_{\text{NL}}^2}   \ , \quad e_{12}^{\rm St} =    -\frac{2 f k_2 \mu _2 \left(k_1 \mu _1+k_2 \mu _2\right)}{k_{\text{NL}}^2}  \nonumber  \ ,   \\
& e_{13}^{\rm St} =   \frac{1}{4 k_1^2 k_3^2 k_{\text{NL}}^2} f \Big( k_1^2 \left(k_1^2-k_2^2+k_3^2\right){}^2 \mu _1^2+2 k_1 k_2 \left(k_1^2-k_2^2+k_3^2\right){}^2 \mu _1 \mu
   _2    \nonumber  \\
   & \hspace{1in} +\left(\left(k_2^2+k_3^2\right) k_1^4-2 \left(k_2^2-k_3^2\right){}^2 k_1^2+\left(k_2^2-k_3^2\right){}^2 \left(k_2^2+k_3^2\right)\right) \mu
   _2^2  \Big)  \ .       \nonumber 
\end{align}
All of the above $e_i^{\rm St}$ are symmetric when swapping $\kvec_2$ and $\kvec_3$, as expected from \eqn{btilde321}.  To see it, one must swap $k_2 \leftrightarrow k_3$ and  $\mu_2 \leftrightarrow \mu_3$, and then replace $\mu_3 = - k_3^{-1}( k_1 \mu_1 + k_2 \mu_2)$.

%%%%%%%%%%%%%%%%
%

%

\section{Binning formula details}\label{app:binningd}
In this appendix, we want to show that the binning formula for the bispectrum 
\begin{equation} \label{firstbinform}
  B_{(l, i), \rm bin}^{r,h}(k_1, k_2, k_3) = \frac{2l+1}{V_T} \left(\prod_i \int_{V_i} \frac{\rmd^3 q_i}{(2 \pi)^3} \right)(2 \pi)^3 \delta_D^{(3)}(\qvec_1+\qvec_2+\qvec_3) \mathcal{P}_l(\mu_i) B^{r,h} (\qvec_1, \qvec_2, \qvec_3)  \ , 
\end{equation}
is equivalent  to
\begin{equation}
    B_{(l, i) , \rm bin}^{r,h} (k_1, k_2, k_3) = \frac{1}{V_T}  \left(\prod_{i} \int_{k_i }  \rmd q_i \, q_i \right) \,  \frac{\beta\(\Delta_q\)}{8 \pi^4} B^{r,h}_{{(l,i)}}(q_1, q_2, q_3) \,  \, ,
    \label{eq:codebin}
\end{equation}
{We do the calculation here for general $l$, which is relevant to us for $l = 0 , 2$.}

Given that the bispectrum is a polynomial in $\mu_1$ and $\mu_2$, and switching to a basis of Legendre polynomials, we can write
\begin{equation}\label{lexpans}
  B^{r,h}(\qvec_1, \qvec_2, \qvec_3) = \sum_{n_1, n_2} B^{r,h}_{n_1, n_2}(q_1, q_2, q_3) \calP_{n_1}(\mu_1) \calP_{n_2}(\mu_2) \, .
\end{equation}
We can focus on the case of $\calP_l(\mu_1)$, since the other cases just correspond to a permutation of $\mu_i$ in \eqn{lexpans}. Let us then start by writing the delta function as an integral over plane waves:
\begin{align}
\begin{split} \label{bin3}
  B^{r,h}_{ ( l, 1) , \rm bin} &  = \frac{2l+1}{V_T} \left(\prod_i \int_{V_i} \frac{\rmd^3 q_i}{(2 \pi)^3}\right) \int \rmd^3 x \, e^{i \xvec \cdot ( \qvec_1 + \qvec_2 + \qvec_3)} \mathcal{P}_l(\mu_1) \\
  &  \hspace{2in}  \times \sum_{n_1, n_2} B^{r,h}_{n_1,n_2}(q_1, q_2, q_3) \calP_{n_1}(\mu_1) \calP_{n_2}(\mu_2) \, .
\end{split}
\end{align}
The integral  over $\rmd^2 \hat{q}_3$, is just
\begin{equation} \label{bessel0}
  \int \rmd^2 \hat{q}_3 \, e^{i \vec{q}_3 \cdot \vec x} = 4 \pi j_0(q_3 x) \, .
\end{equation}
And the rest of the exponentials we expand the plane wave:
\begin{equation}
  e^{i \qvec_1 \cdot \vec x} = \sum_{l_1} i^{l_1} (2 l_1 + 1) j_{l_1}(q_1 x) \calP_{l_1}(\hat q_1 \cdot \hat x) .
\end{equation}
Putting all of this into \eqn{bin3} we get: 
\begin{equation}
  \begin{split}
    B^{r,h}_{ ( l, 1) , \rm bin} &= \frac{2l+1}{V_T} \sum_{n_1, n_2} \sum_{l_1,l_2}i^{l_1+l_2} (2 l_1 + 1) (2 l_2 + 1) \left( \prod_{i=1}^2 \int_{V_i} \frac{\rmd^3 q_i}{(2 \pi)^3} \right) \int_{k_3  } \frac{\rmd q_3}{2 \pi^2} q_3^2 B^{r,h}_{n_1,n_2}(q_1, q_2, q_3) \\
    &\times\int \rmd^3 x   j_0(q_3 x)  j_{l_1}(q_1 x) j_{l_2}(q_2 x) \calP_{l_1}(\hat q_1 \cdot \hat x) \calP_{l_2}(\hat q_2 \cdot \hat x) \calP_l(\hat q_1 \cdot \hat z) \calP_{n_1}(\hat q_1 \cdot \hat z) \calP_{n_2}(\hat q_2 \cdot \hat z)\, .
  \end{split}
\end{equation}
Now we can do all the angular integrals, using the formula,
\begin{equation}
  \int \rmd^2 \hat x \calP_{l_1}(\hat q_1 \cdot \hat x) \calP_{l_2}(\hat q_2 \cdot \hat x) = \delta_{l_1,l_2} \,  \frac{4 \pi }{2 l_1 + 1} \calP_{l_1}(\hat q_1 \cdot \hat q_2)  \ , 
\end{equation}
to evaluate
\begin{equation}
  \begin{split}
  \int \rmd^2 \hat x \int \rmd^2 & \hat q_1 \int \rmd^2  \hat q_2 \, \calP_{l_1}(\hat q_1 \cdot \hat x) \calP_{l_2}(\hat q_2 \cdot \hat x) \calP_l(\hat q_1 \cdot \hat z) \calP_{n_1}(\hat q_1 \cdot \hat z) \calP_{n_2}(\hat q_2 \cdot \hat z) \\
  &= \delta_{l_1,l_2} \,  \frac{4 \pi }{2 l_1 + 1}  \int \rmd^2 \hat q_1 \int \rmd^2 \hat q_2 \, \calP_{l_1}(\hat q_1 \cdot \hat q_2) \calP_l(\hat q_1 \cdot \hat z) \calP_{n_1}(\hat q_1 \cdot \hat z) \calP_{n_2}(\hat q_2 \cdot \hat z) \\
  &= \delta_{l_1,l_2} \, \delta_{l_1,n_2} \, \frac{(4 \pi)^2}{(2 l_1 + 1)^2}  \int \rmd^2 \hat q_1 \, \calP_{l_1}(\hat q_1 \cdot \hat z) \calP_l(\hat q_1 \cdot \hat z) \calP_{n_1}(\hat q_1 \cdot \hat z) \\
  &= \delta_{l_1,l_2} \, \delta_{l_1,n_2} \, \frac{(4 \pi)^3}{(2 l_1 + 1)^2}
  \begin{pmatrix}
    n_2 & l & n_1 \\
    0 & 0 & 0
  \end{pmatrix}^2 \ , 
  \end{split} 
\end{equation}
and additionally we used the integral of three Legendre polynomials in terms of the Wigner 3-j symbol in the last line.
We are now only left with integrals over the magnitudes: 
\begin{align}
\begin{split}
    B^{r,h}_{ ( l, 1 ), \rm bin} & = 4 \pi \frac{(2l + 1)}{V_T} \sum_{n_1, n_2}(-1)^{n_2} \( \prod_{i=1}^3 \int_{k_i 
} \frac{\rmd q_i}{2 \pi^2} q_i^2 \) \\
&  \quad   \times \int_0^\infty \rmd x x^2 j_{n_2}(q_1 x) j_{n_2}(q_2 x)  j_0(q_3 x)
    \begin{pmatrix}
      n_2 & l & n_1 \\
      0 & 0 & 0
    \end{pmatrix}^2 B^{r,h}_{n_1,n_2}(q_1, q_2, q_3)  \, .
\end{split}
\end{align}
Further using the following integral over three spherical Bessel functions:
\begin{equation} \label{threebesselint1}
  \int_0^\infty \rmd x x^2 j_{n_2}(q_1 x) j_{n_2}(q_2 x)  j_0(q_3 x) = \frac{\pi}{4 q_1 q_2 q_3} \beta(\hat{q}_1 \cdot \hat{q}_2) \calP_{n_2}\(\frac{q_1^2 + q_2^2 - q_3^2}{2 q_1 q_2}\)  \ , 
\end{equation}
where $\beta ( \Delta ) = 1$ for $-1 < \Delta < 1$, $\beta(\Delta ) = 1/ 2 $ for $\Delta = \pm 1$, and $\beta(\Delta ) = 0 $ otherwise, and recognizing that the last Legendre is $\calP_{n_2}(- \hat q_1 \cdot \hat q_2) = (-1)^{n_2} \calP_{n_2}(\hat q_1 \cdot \hat q_2)$, we can put everything together and 
we get that \eqn{firstbinform} reduces to   
\begin{equation}
  B^{r,h}_{ ( l, 1), \rm bin} = \frac{2l+1}{V_T} \sum_{n_1, n_2} \left(\prod_{i=1}^3 \int_{k_i }\rmd q_i q_i \right)\frac{\beta(\hat{q}_1 \cdot \hat{q}_2)}{8 \pi^4} \calP_{n_2}(\hat{q}_1 \cdot \hat{q}_2) 
  \begin{pmatrix}
    n_2 & l & n_1 \\
    0 & 0 & 0
  \end{pmatrix}^2 B^{r,h}_{n_1,n_2}(q_1, q_2, q_3)  \, .
\end{equation}

To get our formula~\eqref{eq:codebin}, it is now sufficient to show that the unbinned bispectrum satisfies
\begin{equation}
  B^{r,h}_{( l, 1) }(q_1, q_2, q_3) = (2l+1)\sum_{n_1, n_2} \calP_{n_2}(\hat{q}_1 \cdot \hat{q}_2) 
  \begin{pmatrix}
    n_2 & l & n_1 \\
    0 & 0 & 0
  \end{pmatrix}^2 B^{r,h}_{n_1,n_2}(q_1, q_2, q_3) \, .
\end{equation}
So next, we write the left hand side of the above explicitly, and expand the redshift space bispectrum, plugging \eqn{lexpans} into {a generalization of \eqn{monopoleeq} and \eqn{bispmultipolesused}}: 
\begin{equation} 
  B^{r,h}_{(l, 1)}(q_1, q_2, q_3) =(2l+1) \sum_{n_1, n_2} \int_{-1}^1 \frac{\rmd \mu_1}{2} \int_0^{2 \pi} \frac{\rmd \phi}{2 \pi} \mathcal{P}_l(\mu_1) \calP_{n_1}(\mu_1) \calP_{n_2}(\mu_2) B^{r,h}_{n_1, n_2}(q_1, q_2, q_3)  \, .
\end{equation}
This can be calculated in a coordinate system in which we fix $\hat{q}_1$, $\hat{q}_2$ and integrate over $\rmd^2 \hat z$:
\begin{equation} \label{bispectintegra}
  B^{r,h}_{( l, 1 ) }(q_1, q_2, q_3) = (2l+1)\sum_{n_1, n_2} \int \frac{\rmd^2 \hat z}{4 \pi} \calP_l(\hat q_1 \cdot \hat z) \calP_{n_1}(\hat q_1 \cdot \hat z) \calP_{n_2}(\hat q_2 \cdot \hat z) B^{r,h}_{n_1, n_2}(q_1, q_2, q_3)  \, .
\end{equation}
Next, we use that the product of two Legendre polynomials is
\begin{equation} \label{twolegexpand}
  \calP_l(\hat q_1 \cdot \hat z) \calP_{n_1}(\hat q_1 \cdot \hat z) = \sum_{L=|n_1-l|}^{n_1+l} (2 L + 1) \begin{pmatrix}
    n_1 & l & L \\
    0 & 0 & 0
  \end{pmatrix}^2 \calP_{L}(\hat q_1 \cdot \hat z) \ , 
\end{equation}
and plug \eqn{twolegexpand} into \eqn{bispectintegra} to get  
\begin{align}
  \begin{split}
  B^{r,h}_{( l, 1 )}(q_1, q_2, q_3) &= (2l+1)\sum_{n_1, n_2} \sum_{L=|n_1-l|}^{n_1+l} (2 L + 1) \begin{pmatrix}
    n_1 & l & L \\
    0 & 0 & 0
  \end{pmatrix}^2 \\
  & \hspace{1.5in} \times  \int \frac{\rmd^2 \hat z}{4 \pi} \calP_{L}(\hat q_1 \cdot \hat z) \calP_{n_2}(\hat q_2 \cdot \hat z) B^{r,h}_{n_1, n_2}(q_1, q_2, q_3) \\
  &= (2l+1)\sum_{n_1, n_2} \begin{pmatrix}
    n_1 & l & L \\
    0 & 0 & 0
  \end{pmatrix}^2 \calP_{n_2}(\hat q_1 \cdot \hat q_2)  B^{r,h}_{n_1, n_2}(q_1, q_2, q_3) \ , 
\end{split}
\end{align}
as desired.

For completeness we also calculate the volume
\begin{equation}
  \begin{split}
  V_T &= \left(\prod_i \int_{V_i} \frac{\rmd^3 q_i}{(2 \pi)^3} \right)(2 \pi)^3 \delta_D^{(3)}(\qvec_1+\qvec_2+\qvec_3)  = \left(\prod_i \int_{V_i} \frac{\rmd^3 q_i}{(2 \pi)^3}\right) \int \rmd^3 x \, e^{i \qvec_1 \cdot \vec x} e^{i \qvec_2 \cdot \vec x} e^{i \qvec_3 \cdot \vec x} \, .
  \end{split}
\end{equation}
We then integrate over the plane waves using \eqn{bessel0} and the three Bessel functions using \eqn{threebesselint1} to get 
\begin{equation}
  \begin{split}
  V_T & =  \left(\prod_i \int_{k_i} \rmd q_i \,  q_i \right)\frac{\beta(\hat{q}_1 \cdot \hat{q}_2)}{8 \pi^4} \ . 
  \end{split}
\end{equation}

 \begin{figure}[h!]
  \centering
  \includegraphics[width=0.99\textwidth]{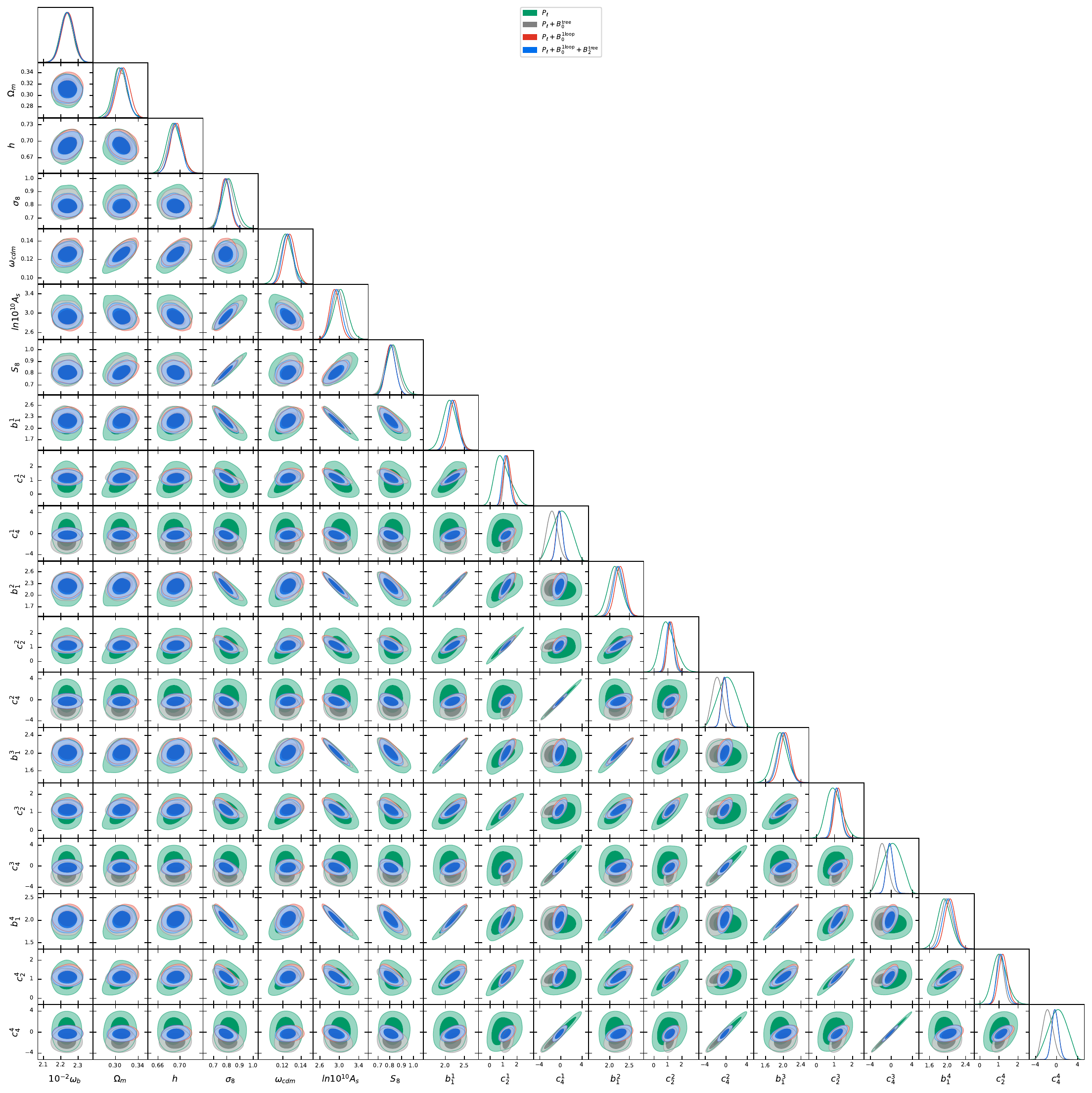}
      \caption{\footnotesize  
   Full triangle plots from the analysis of BOSS power spectrum multipoles $P_\ell$ at one loop, bispectrum monopole $B_0$ at {tree level or} one loop, and bispectrum quadrupole $B_2$ at tree level. 
   }
  \label{fig:tri_full}
  \end{figure}
 
%%%%%%%%%%%%%%%
%
%

\section{Additional parameter posteriors}\label{app:constraints}

In fig.~\ref{fig:tri_full}, we show the full triangle plots obtained fitting BOSS 4 skies $P_\ell + B_0 + B_2$. 
In tab.~\ref{tab:eftparams}, we show the 68\%-credible intervals of $b_1, c_2$, and $c_4$ obtained on this same fit.  

\begin{table}[h!]
\centering
\scriptsize
\begin{tabular}{|ll|c|c|c|c|}
\hline
\multicolumn{2}{|c|}{mean $\pm \sigma$}                  & $b_1$          & $c_2$                  & $c_4$                   \\ \hline
\multirow{3}{*}{CMASS NGC}               &
 $P_\ell$         & $2.12\pm 0.18$ & $0.95^{+0.45}_{-0.68}$ & $0.2\pm 1.7$            \\ 
                           & $P_\ell+B_0$     & $2.23\pm 0.13$ & $1.27\pm 0.26$         & $-0.29^{+0.55}_{-0.61}$ \\ 
                           & $P_\ell+B_0+B_2$ & $2.19\pm 0.13$ & $1.18^{+0.22}_{-0.28}$ & $-0.25^{+0.54}_{-0.60}$ \\ \hline
\multirow{3}{*}{CMASS SGC}               &
 $P_\ell$         & $2.13\pm 0.18$ & $1.01^{+0.45}_{-0.62}$ & $0.2\pm 1.7$            \\ 
                           & $P_\ell+B_0$     & $2.27\pm 0.13$ & $1.23\pm 0.26$         & $-0.32^{+0.56}_{-0.63}$ \\
                           & $P_\ell+B_0+B_2$ & $2.22\pm 0.14$ & $1.14^{+0.23}_{-0.27}$ & $-0.27\pm 0.60$         \\ \hline
\multirow{3}{*}{LOWZ NGC}               &
 $P_\ell$         & $1.93\pm 0.16$ & $0.98^{+0.37}_{-0.47}$ & $0.2\pm 1.7$            \\ 
                           & $P_\ell+B_0$     & $2.04\pm 0.12$ & $1.23^{+0.21}_{-0.24}$ & $-0.26\pm 0.64$         \\ 
                           & $P_\ell+B_0+B_2$ & $2.00\pm 0.12$ & $1.14^{+0.20}_{-0.24}$ & $-0.27^{+0.60}_{-0.67}$ \\ \hline
\multirow{3}{*}{LOWZ SGC}               &
 $P_\ell$         & $1.93\pm 0.15$ & $1.04^{+0.34}_{-0.40}$ & $0.2\pm 1.7$            \\ 
                           & $P_\ell+B_0$     & $2.05\pm 0.12$ & $1.21^{+0.21}_{-0.24}$ & $-0.30\pm 0.65$         \\ 
                           & $P_\ell+B_0+B_2$ & $2.02\pm 0.12$ & $1.12^{+0.20}_{-0.24}$ & $-0.30^{+0.61}_{-0.68}$ \\ \hline
\end{tabular}
\caption{\footnotesize 68\%-credible intervals of $b_1, c_2$, and $c_4$ from the analysis of BOSS power spectrum multipoles $P_\ell$ at the one-loop, bispectrum monopole $B_0$ at the one-loop, and bispectrum quadrupole $B_2$ at tree-level.  }
\label{tab:eftparams}
\end{table}

%%%%%%%%%%%%%%%%
%
%

\pagebreak
\bibliographystyle{JHEP}
\small
\bibliography{references_3}

\end{document}